\documentclass[aps,prl,preprint,groupedaddress,showpacs]{revtex4-1}
\usepackage{graphicx}
\usepackage{amsmath}
\begin{document}
\title{Torus-Knotted Electromagnetic Fields}
\author{Manuel Array\'as and Jos\'e L. Trueba}
\affiliation{\'Area de Electromagnetismo, Universidad Rey Juan
Carlos, Camino del Molino s/n, 28943 Fuenlabrada, Madrid, Spain}

\begin{abstract}
  We present a new range of solutions of Maxwell's equations in
  vacuum in which the topology of the field lines is that of the whole
  torus knots set. Knotted electromagnetic fields are solutions of the
  Maxwell equations in vacuum in which magnetic lines, and also
  electric lines, have some kind of linkage. These solutions may play
  an important role in fundamental physics problems but also have
  practical interest and applications. 
\end{abstract}

\date{\today}
\maketitle

\section{Introduction}
Since Maxwell's conjecture that light waves are electromagnetic waves, and the long struggle 
to gain general acceptance, it seems that electromagnetic theory of light has obtained 
its limits of serviceability: capable of explaining the main features of all phenomena 
concerning with the propagation of light, but failure to elucidate processes in which the 
finer features of the interaction between matter and field are manifested \cite{Born}. 
However, the fundamental nature of the difficulties that the electromagnetic theory of 
light finds has always paved the road for new discoveries. One example is the first 
attempts by Hertz to generalise Maxwell's laws to moving bodies. Maxwell's equations 
remains a fundamental cornerstone of our undertanding of nature.

In this paper, we present a new class of solutions of Maxwell's
equations in vacuum with non-trivial topology. To be specific: with
torus knot topology. One particular solution of this type was found
based on the Hopf fibration \cite{Trautman,Ran89,Ran95,Ran97,Irv08}. It had
some interesting properties, like the linkage of the field lines
preserved in time. Recently some progress has been made in the
understanding of the interaction of this field with charged
particles \cite{Arr10}. The new class of solutions presented in this
work covers the topology of the whole torus knots set and has previous
solutions as particular cases. By having the topology of the torus
knots set we mean that initially all the magnetic lines and all the
electric lines are linked and closed on the surface of a torus and,
moreover, when time evolves we can find numerically field lines
knotted as a torus knot. These configurations could be important
theoretically, as the stability of electromagnetic fields \cite{Ran96}
may play a role in particle theory \cite{Kelvin,Wheeler,Jehle1,Jehle2}
or even in certain asymptotic limits of string theory \cite{Munoz}.

There are examples in nature where the magnetic fields created by
planets and starts present also toroidal structure and nontrivial
topology of the field lines \cite{Innes}. In the laboratory we find
electromagnetic fields in toroidal geometries for plasma confinement
\cite{Taylor} and nuclear magnetic resonance devices \cite{Jackson}.
So the solutions presented in which follows might be of practical
interest.

\section{Construction of a class of electromagnetic fields with knotted field lines} 
\begin{figure}
\centering
\includegraphics[width=0.25\textwidth]{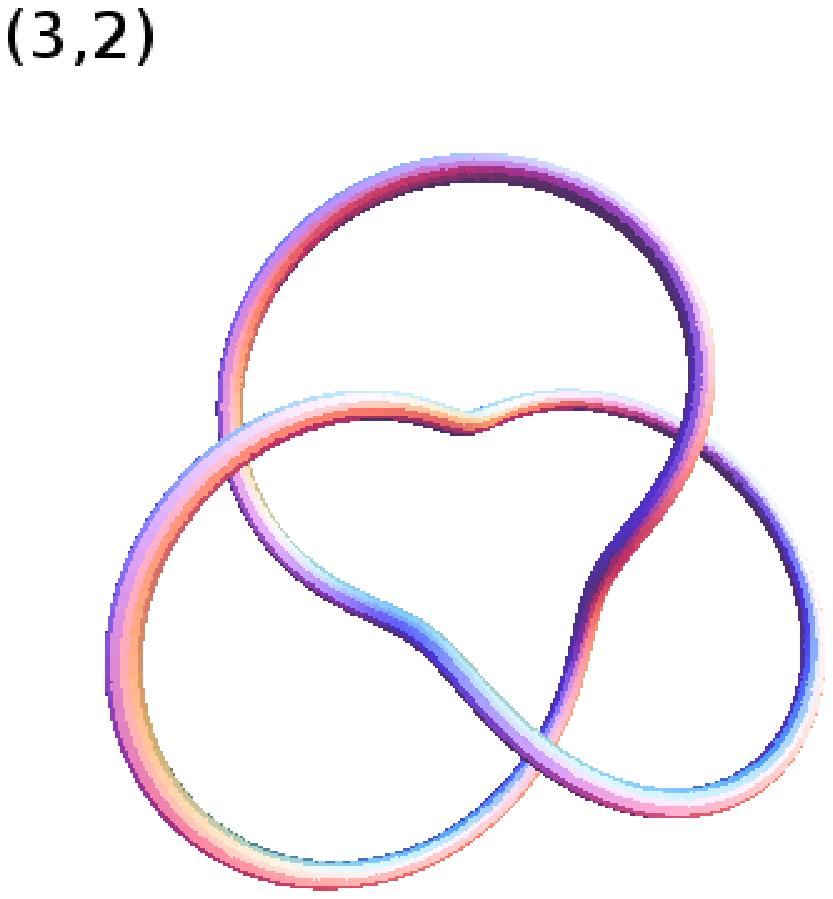}
\includegraphics[width=0.25\textwidth]{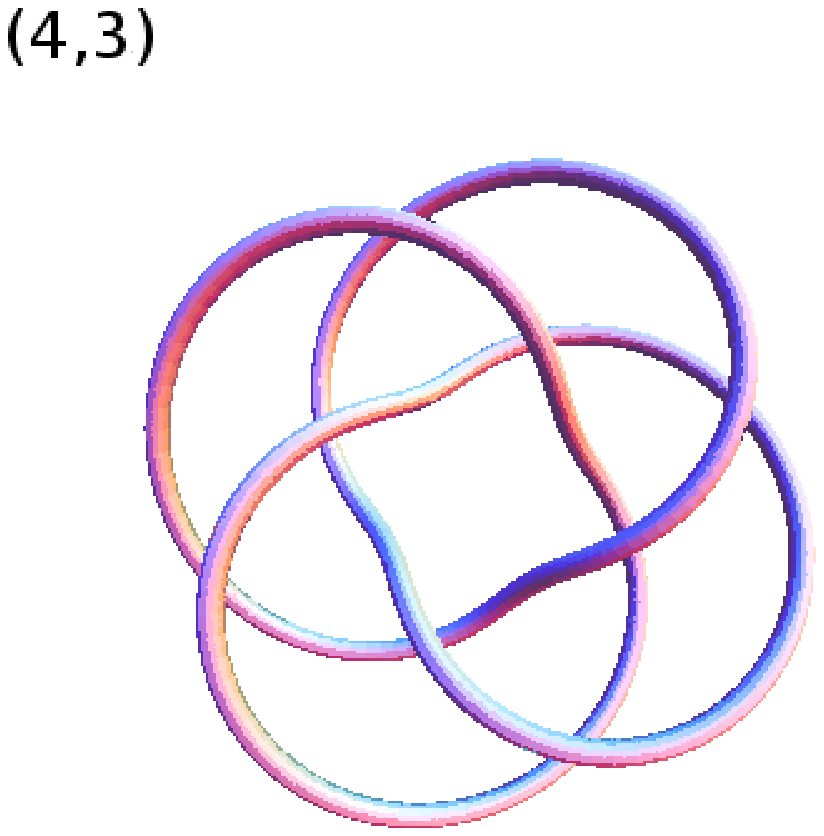}
\includegraphics[width=0.25\textwidth]{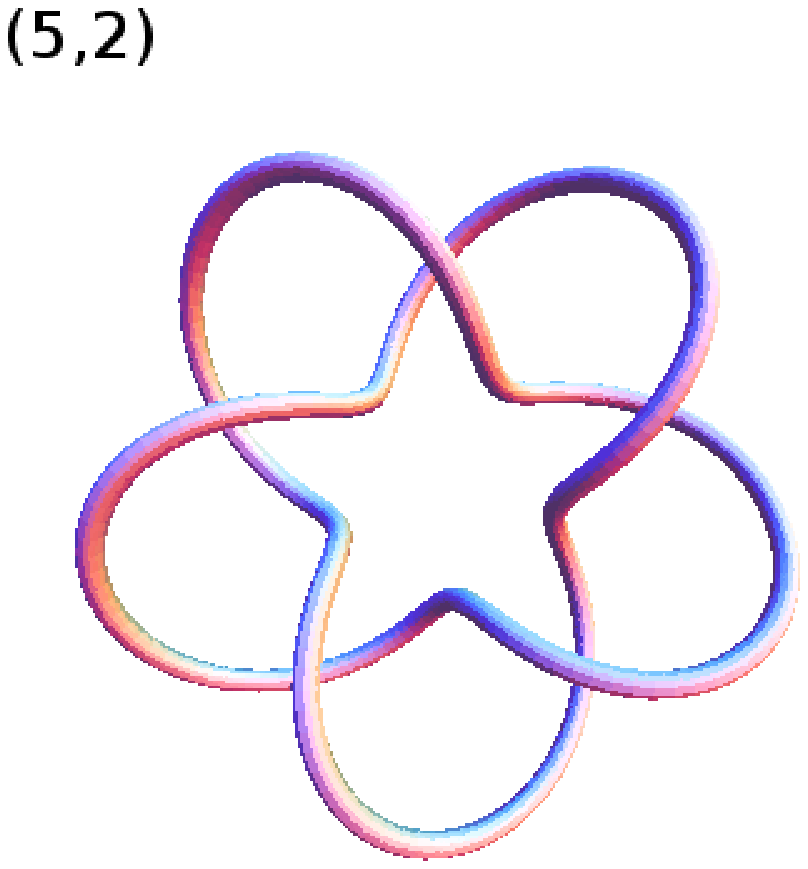}
\includegraphics[width=0.25\textwidth]{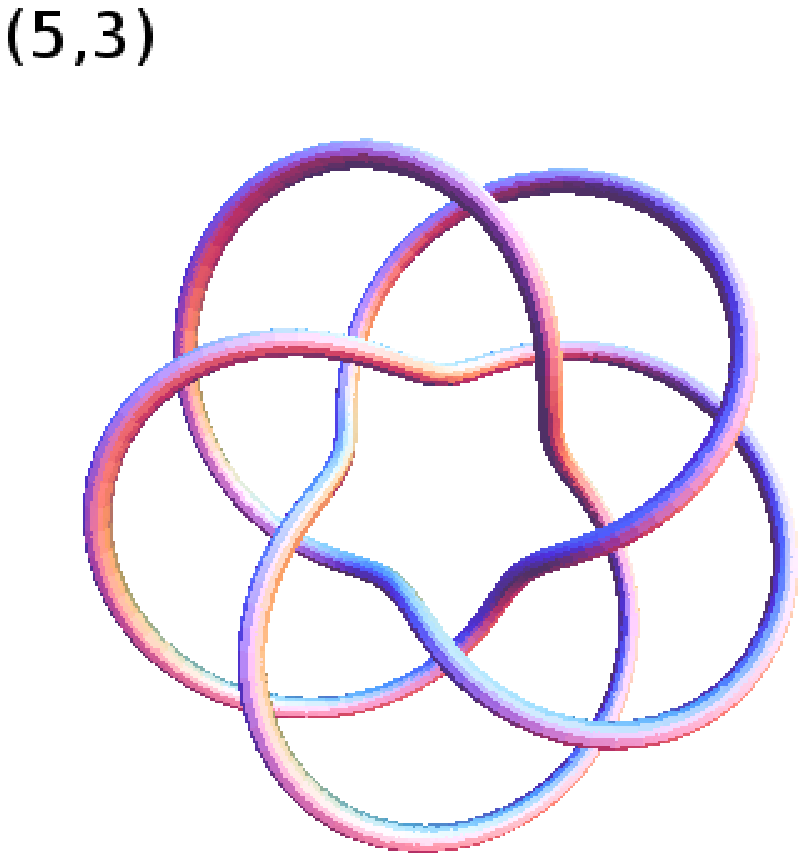}
\includegraphics[width=0.25\textwidth]{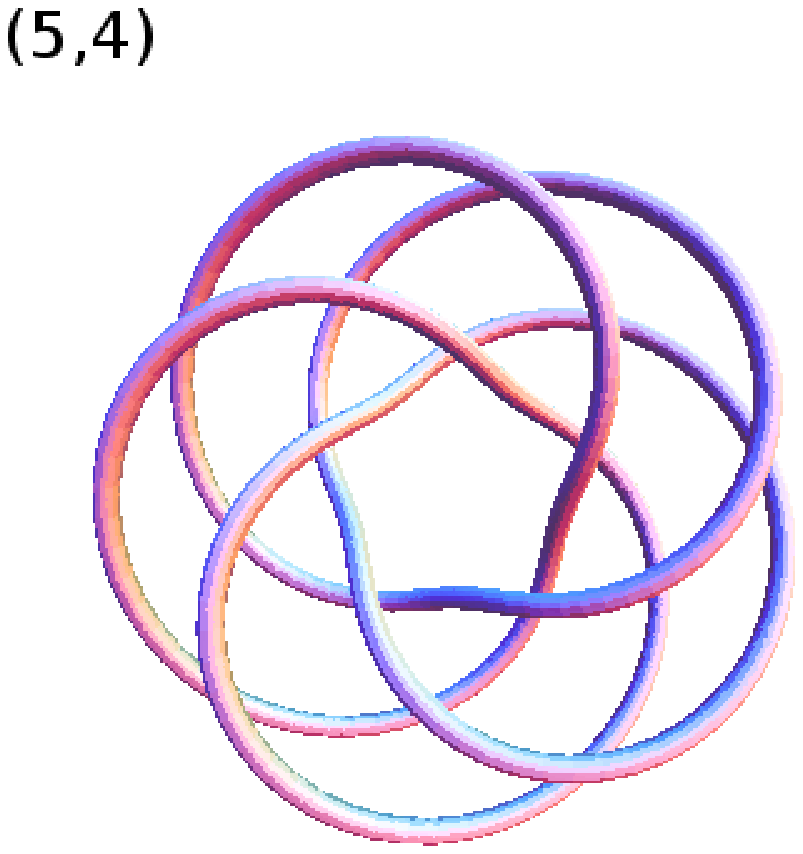}
\includegraphics[width=0.25\textwidth]{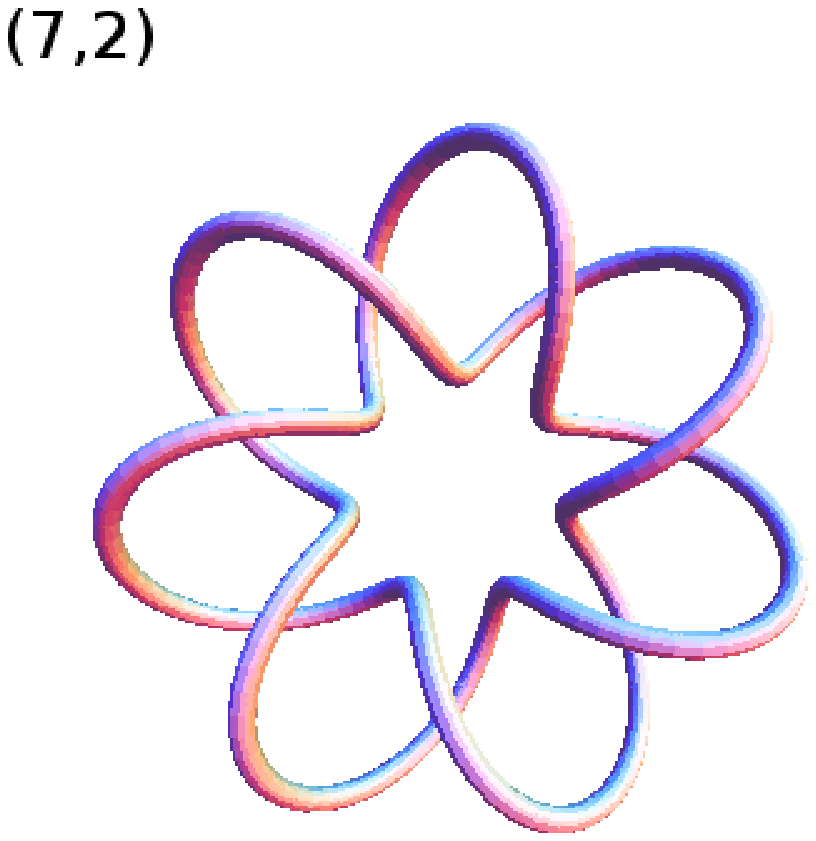}
\includegraphics[width=0.25\textwidth]{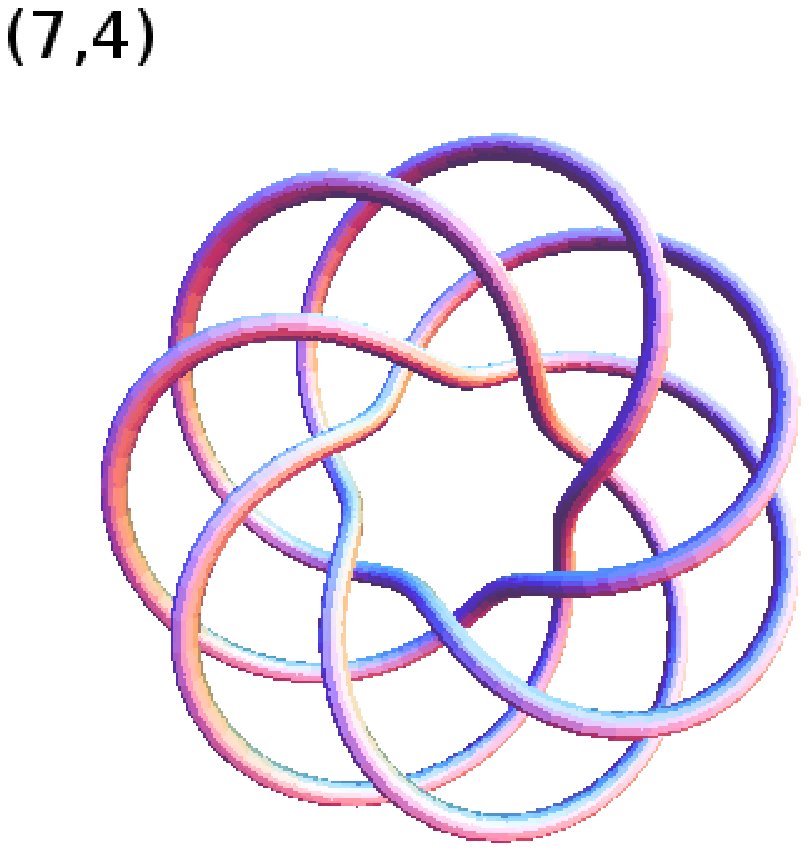}
\includegraphics[width=0.25\textwidth]{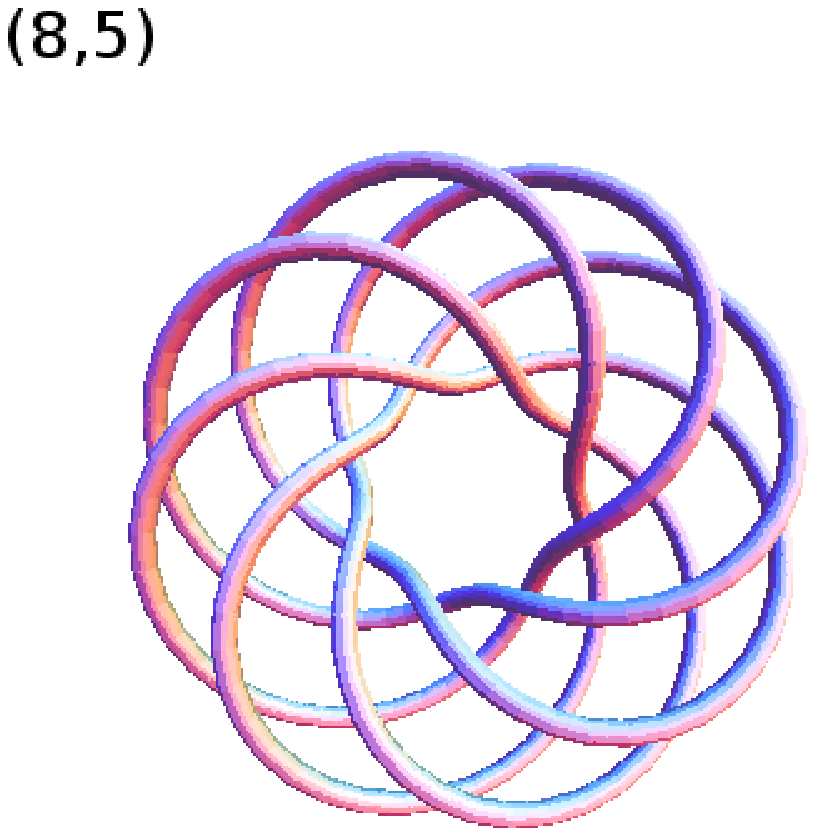}
\includegraphics[width=0.25\textwidth]{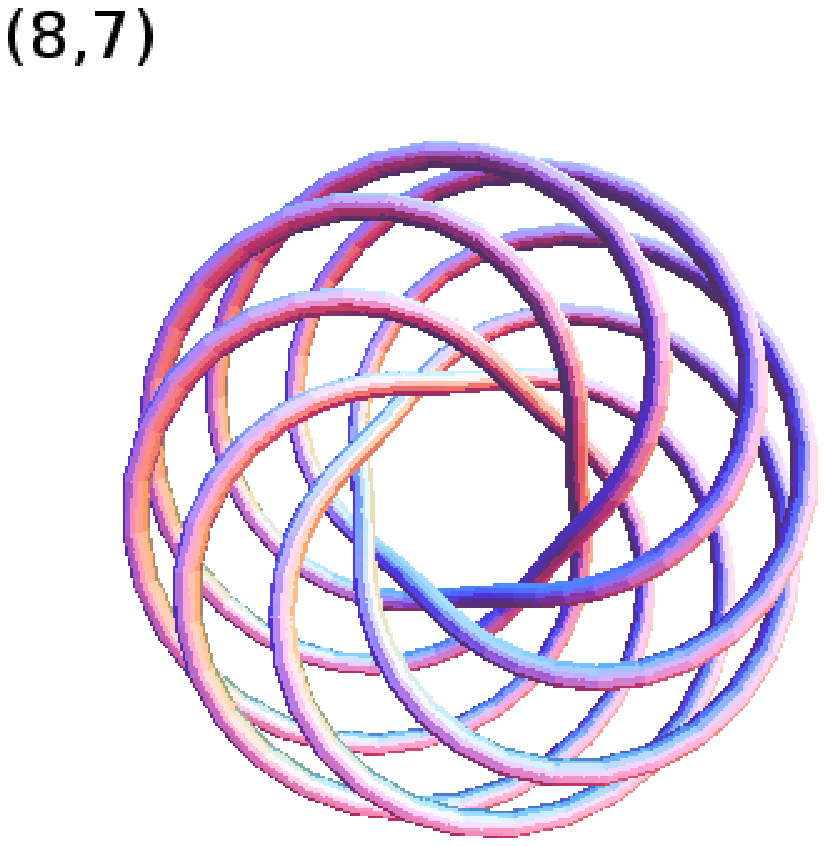}
\caption{Some $(n,m)$ torus knots. These knotted curves are lying on the 
surface of a torus in the three-dimensional space $R^3$. In the figures, 
all the torus knots are called nontrivial. If $n$ or $m$ were equal to 1, 
the knot is said to be trivial since its representation is a circle, that 
can be linked to other ones.} 
\label{atlas}
\end{figure}

Torus knots are knotted curves lying on the surface of a torus. Any
torus knot is defined by two coprime integer numbers $(n,m)$ in such a
way that the curve winds $n$ times around a circle inside the torus
and $m$ times around a line through the hole in the torus. In figure
\ref{atlas} we plot some particular examples.

Here is a method to find some knotted electromagnetic fields in vacuum
(see \cite{Ran01}), i. e. exact solutions of Maxwell's equations
in empty space in which the magnetic lines and/or the electric lines
are knotted curves.
 
Let $\phi ({\bf r})$, $\theta ({\bf r})$, two complex scalar fields
that depend on the position ${\bf r}$ in the physical space $R^3$. In
the cases that we are going to study in this work, both scalar fields
can be considered as maps $\phi, \theta : S^3 \rightarrow S^2$ after
identifying the physical space $R^3$ with $S^3$ (doing so, it is
assumed that both scalars have only one value at infinity) and the
complex plane with $S^2$. These identifications can be done using
stereographic projections and, as we will see later, have important 
consequences in the solutions of Maxwell's equations that we are going 
to obtain. However, a similar 
formalism could be applied to other possibilities in which both scalars 
cannot be considered as maps from $S^3$ to $S^2$.  For example, this 
formalism could be used to construct electromagnetic fields in vacuum 
from the scalars considered in \cite{Dennis} to implement optical 
beams.

If $\phi$ (or $\theta$) is a map from $S^3$ to $S^2$, then the
preimage of any point in $S^2$ is a closed curve in $S^3$. Moreover,
the linking number of a pair of curves in $S^3$ (that are preimages of
two distinct points in $S^2$) is the same for all the pairs of curves
(except at most for a zero-measure set, a property common to torus
knots). The linking number of each pair of curves is called the Hopf 
invariant $H(\phi)$ of the map $\phi$ and it is an integer number.

We will impose that, at some initial time $t=0$, the level curves of
the complex scalar fields $(\phi, \theta)$ coincide with the magnetic 
and electric lines respectively, each one of these lines being labelled 
by the constant value of the corresponding scalar. This can be simply done
by constructing the magnetic and electric fields at $t=0$ as 
\begin{eqnarray} 
{\bf B}({\bf r},0) &=& \frac{\sqrt{a}}{2\pi i}\frac{ \nabla \phi \times \nabla \bar{\phi}}{(1+ \bar{\phi} \phi )^{2}},  \label{knot1} \\
{\bf E}({\bf r},0) &=& \frac{\sqrt{a} c}{2\pi i}\frac{\nabla \bar{\theta} \times \nabla \theta}{(1+ \bar{\theta} \theta)^{2}}, \label{knot2}
\end{eqnarray}
where $a$ is a constant introduced so that the magnetic and electric fields have correct dimensions. 
In the International System of Units, that will be used in this work, $a$ can be expressed as a pure number 
times the Planck constant $\hbar$ times the speed of light $c$ times the vacuum permeability $\mu_{0}$. 
This means that the constant $a$ is related to the strength of the  electromagnetic field. The notation $\bar{\phi}$ 
means the complex conjugate of $\phi$ and $i$ is the imaginary unit. Note that, using vector identities, 
the expressions (\ref{knot1})-(\ref{knot2}) are such that they satisfy $\nabla \cdot {\bf B} = 0$ and 
$\nabla \cdot {\bf E} = 0$ for any scalar fields $\phi$ and $\theta$, which is a necessary condition for 
${\bf B}({\bf r},0)$ and ${\bf E}({\bf r},0)$ to be admissible initial values of an electromagnetic 
field in vacuum.

The construction given in equations (\ref{knot1})-(\ref{knot2}), along with the fact that $\phi$ and $\theta$ are 
stereographic projections of maps from $S^3$ to $S^2$, assures that all pairs of lines of the field ${\bf B} ({\bf r},0)$
are linked, and that the linking number is the same for all the pairs and it is given by the Hopf index $H(\phi)$ of the map 
$\phi$. Similarly, all pairs of lines of the field ${\bf E} ({\bf r},0)$ are linked, the linking number of all pairs 
of lines is the same and it is given by the Hopf index $H (\theta)$ of the map $\theta$. Thus,
at $t=0$, the linkage of all the magnetic and the electric lines is set by this construction.

It is convenient to introduce dimensionless coordinates $(X, Y, Z,
T)$, related to the physical ones $(x, y, z, t)$ in the SI of units by
$(X, Y, Z, T) = (x, y, z, c t)/L_{0}$, being $c$ the speed of
light. We will also use the relation $r^2 /L_{0}^2 = (x^2 + y^2
+z^2)/L_{0}^2= X^2 + Y^2 +Z^2 =R^2$, where $L_{0}$ is a constant with
dimensions of length that can be considered to be the characteristic
size of the knot ($L_{0}$ is related to the mean quadratic radius of
the energy distribution of the electromagnetic field \cite{Arr10}). Our
starting point is the following choice of complex scalar fields,
\begin{eqnarray}
\phi &=& \frac{(X+iY)^{(n)}}{(Z+i(R^{2}-1)/2)^{(m)}} , \label{knot3} \\ 
\theta &=& \frac{(Y+iZ)^{(l)}}{(X+i(R^{2}-1)/2)^{(s)}} . \label{knot4}
\end{eqnarray}
where $n$, $m$, $l$ and $s$ are positive integer numbers. These fields are related to the 
Seifert fibrations \cite{Dufraine}. The main difference is that, 
in equations (\ref{knot3})-(\ref{knot4}), the notation $\eta^{(n)}$, $\eta$ being a complex number, means to 
leave the modulus of $\eta$ invariant while the phase of $\eta$ is multiplied by $n$. 

Level curves of the complex scalar field $\phi$ given by expression
(\ref{knot3}) are $(n,m)$ linked torus knots. This means that if we
choose any complex number, say $1+i$, the equation $\phi = 1+i$ gives
a curve in the space $R^3$, and this curve is a $(n,m)$ torus knot. If
we choose any other value, say $4-7i$, the equation $\phi=4-7i$ gives
another $(n,m)$ torus knot. The tangent vectors of such
curves are parallel to $\nabla \operatorname{Re}(\phi)\times\nabla
\operatorname{Im}(\phi)$. Moreover, both curves are linked and their
linking number is, precisely, the Hopf index of $\phi$, that is $H
(\phi) = nm$ in this case.  This occurs for any level curves of
$\phi$. Since level curves of $\phi$ coincide, through equation
(\ref{knot1}), with magnetic lines at $t=0$, we can say that any pair
of magnetic lines at $t=0$ is a linked pair of $(n,m)$ torus knots,
and that the linking number is $nm$. The same can be said about the
complex scalar field $\theta$ given by equation (\ref{knot4}) and the
electric field at $t=0$ given by equation (\ref{knot2}), so that any
pair of electric lines at $t=0$ is a linked pair of $(l,s)$ torus
knots and the linking number is $H (\theta) = l s$.

The magnetic helicity $h_{m}$ of these fields, which is a measure of
the mean value of the linkage of the magnetic lines \cite{Ricca92},
and the electric helicity $h_{e}$, have initial values related,
through equations (\ref{knot1})-(\ref{knot2}), to the Hopf indices of
$\phi$ and $\theta$, respectively. This is only true when (i) the
magnetic and the electric fields have the form given by equations
(\ref{knot1})-(\ref{knot2}), and (ii) the scalar fields $\phi$ and
$\theta$ are stereographic projections of maps from $S^3$ to
$S^2$. Consequently, the initial values of the magnetic and the
electric helicities of these electromagnetic fields are given by
\begin{eqnarray}
h_{m}(t=0)  &=& \frac{1}{2 \mu_{0}} \int d^3 r \, {\bf A} ({\bf r},0) \cdot {\bf B} ({\bf r}, 0) = \frac{a}{2 \mu_{0}} \, H (\phi) , \label{knot5} \\
h_{e}(t=0)  &=& \frac{\varepsilon_{0}}{2} \int d^3 r \, {\bf C} ({\bf r},0) \cdot {\bf E} ({\bf r}, 0) = \frac{a}{2 \mu_{0}} \, H (\theta) , \label{knot6}
\end{eqnarray}
where $\mu_{0}$ is the vacuum permeability, $\varepsilon_{0}$ is the
vacuum permittivity, and ${\bf A}$ and ${\bf C}$ are vector potentials
for the magnetic and electric fields, so that ${\bf B} = \nabla \times
{\bf A}$ and ${\bf E} = \nabla \times {\bf C}$. The constant $a/(2
\mu_{0})$ is a unit of helicity. For $t \neq 0$, equations 
(\ref{knot5})-(\ref{knot6}) are not true. This is due to the
fact that the linkage of the field lines may change during time
evolution (see \cite{Arr11} for an example, in which this kind of
change was presented). However, there is also an important case,
called the Ra\~nada-Hopf electromagnetic knot, in which $n=m=l=s=1$,
where the magnetic and electric helicities take the values given by 
equations (\ref{knot5})-(\ref{knot6}) for any time.

Even if the magnetic and electric helicities may change with time, the electromagnetic helicity
\begin{equation}
h = h_{m} + h_{e},
\label{knot7}
\end{equation}
is a constant of the motion for any electromagnetic field in vacuum. Thus, for the electromagnetic fields given
by equations (\ref{knot1})-(\ref{knot2}) at $t=0$, the electromagnetic helicity satisfies
\begin{equation}
h = \frac{a}{2 \mu_{0}} \, (nm+ls) ,
\label{knot8}
\end{equation}
and this quantity is constant during time evolution since it is conserved by Maxwell's equations 
in vacuum.

\begin{figure}
\centering
\includegraphics[width=0.3\textwidth]{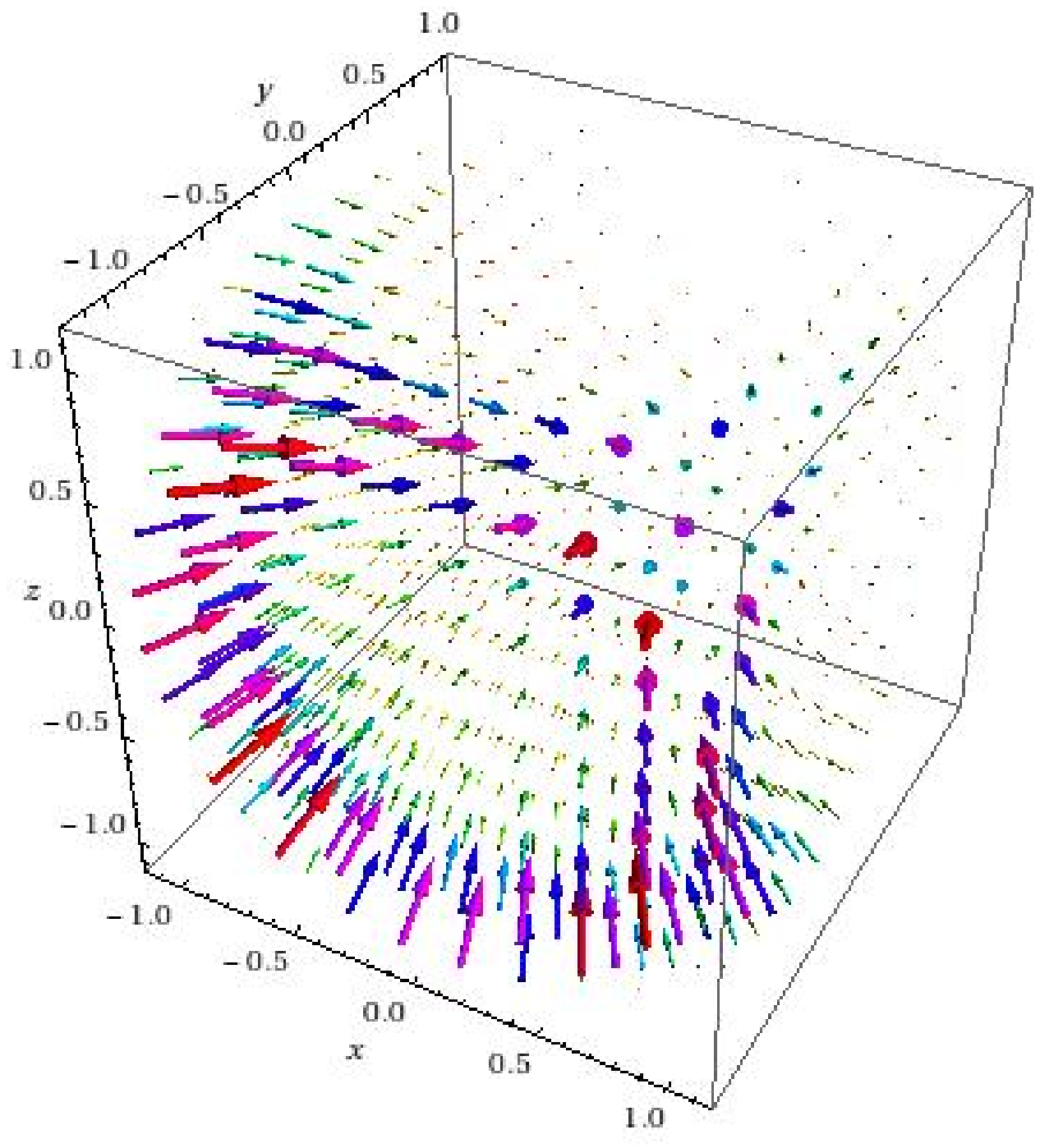}
\includegraphics[width=0.3\textwidth]{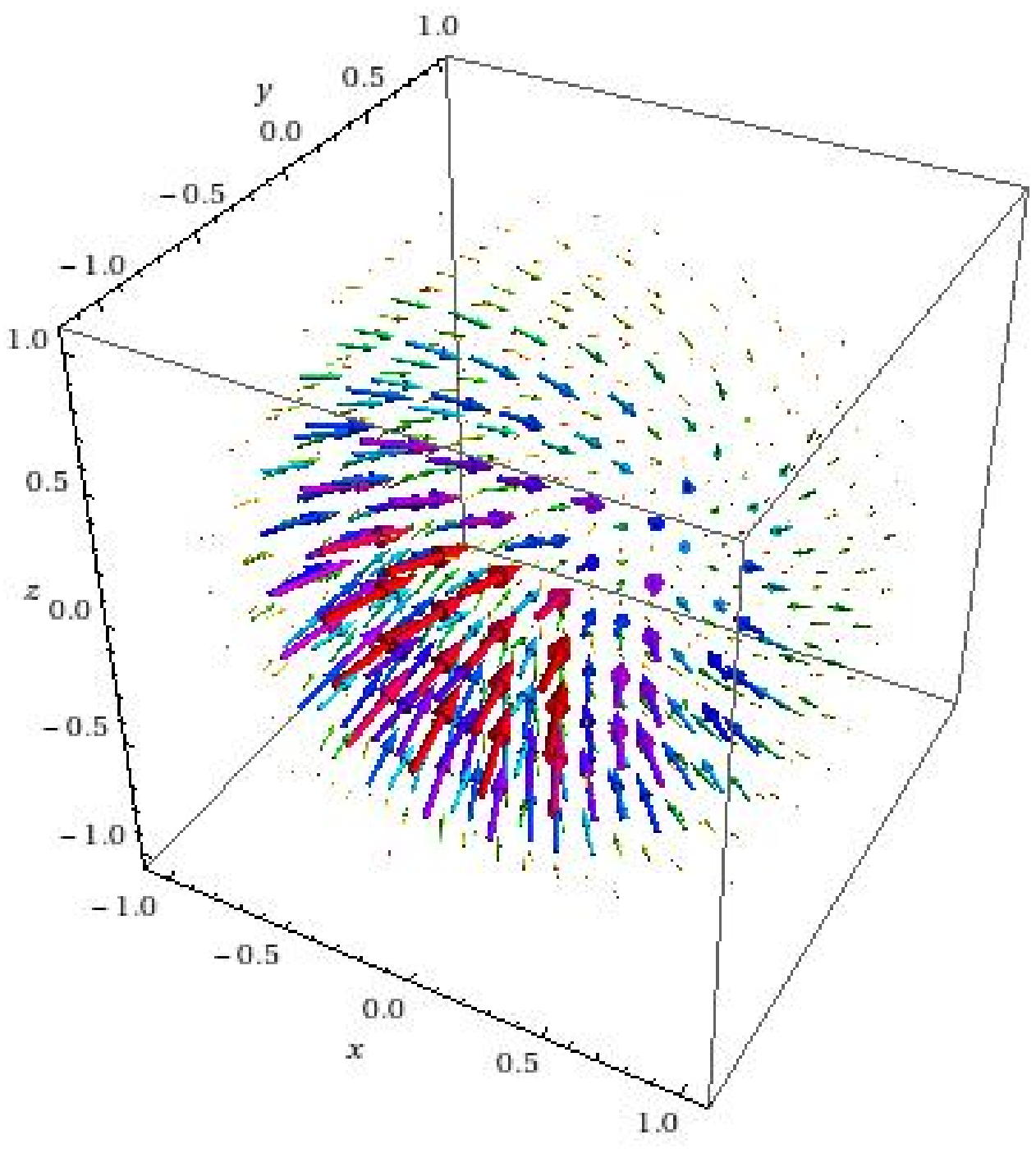}
\includegraphics[width=0.3\textwidth]{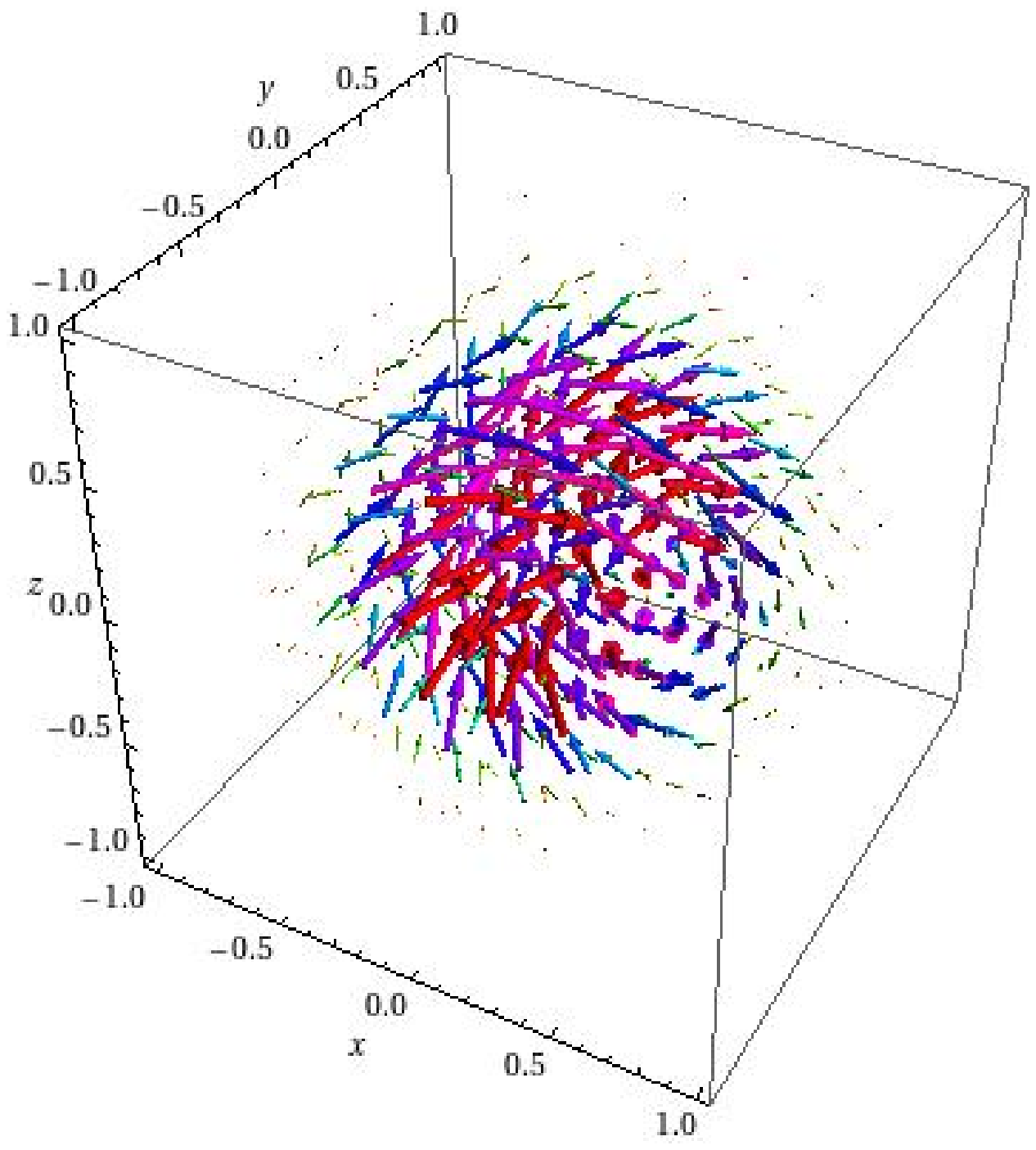}
\includegraphics[width=0.3\textwidth]{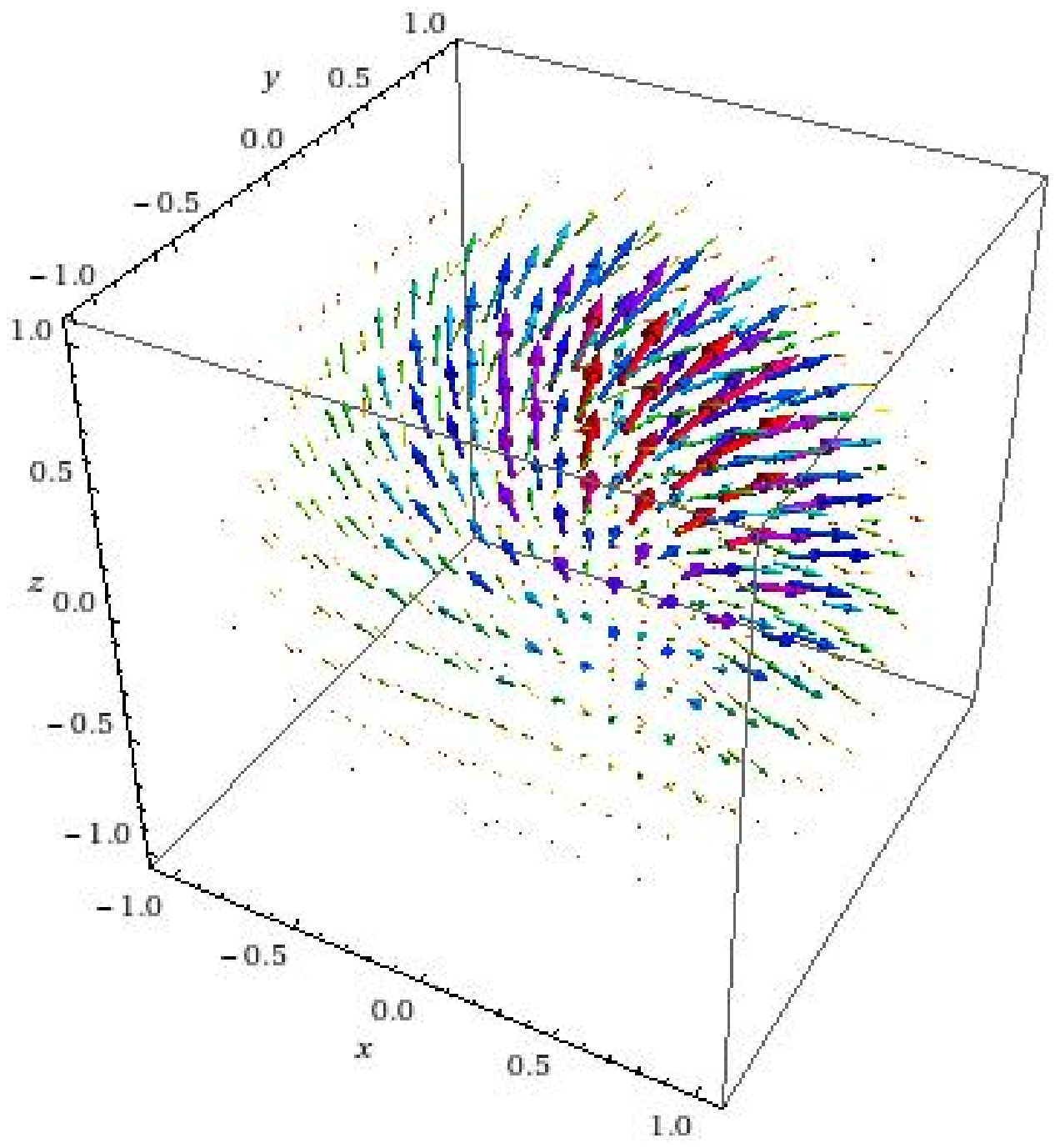}
\includegraphics[width=0.3\textwidth]{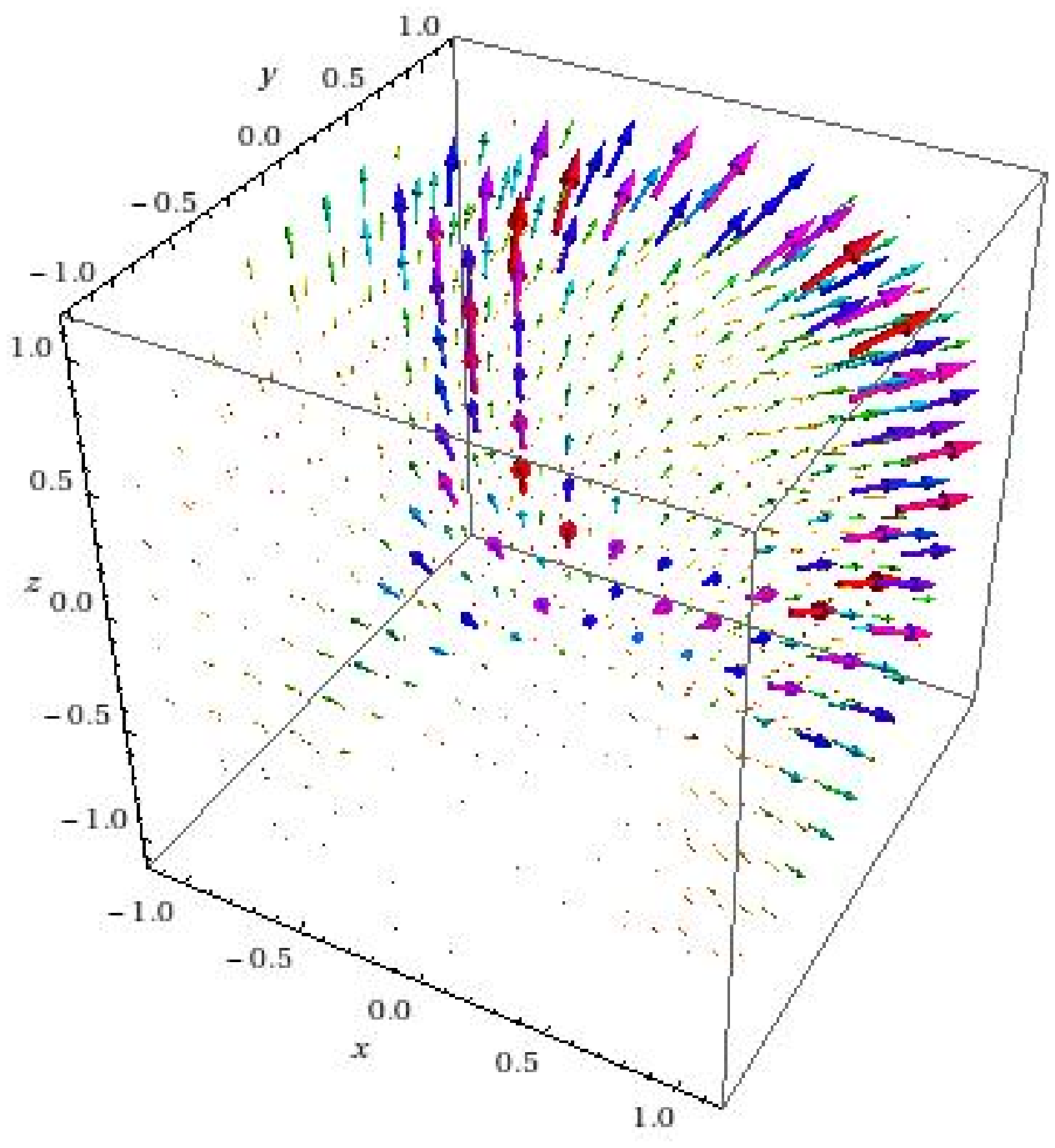}
\caption{Time evolution of the Poynting vector ${\bf E} \times {\bf B} / \mu_{0}$ for the electromagnetic field given by equations (\ref{knot10})-(\ref{knot11}) in the case in which $m=4, n=3, s=3, l=2$. We plot the Poynting vectors at some points of space for $T=-2$, $T=-1$, $T=0$, $T=1$ and $T=2$ (left to right, up to bottom), where $T= c t /L_{0}$. A focusing effect of the energy density flux is observed at $T=0$.} 
\label{poynting}
\end{figure}

To solve Maxwell's equations in vacuum with the initial conditions given by expressions (\ref{knot1})-(\ref{knot2}), Fourier analysis can be used. First one has to compute the Fourier transforms of the initial conditions as
\begin{eqnarray}
{\bf E}_{0} ({\bf k}) &=& \frac{1}{(2 \pi )^{3/2}} \int d^3 r \, {\bf E} ({\bf r}, 0) \, e^{i {\bf k} \cdot {\bf r} } \label{e0} \\
{\bf B}_{0} ({\bf k}) &=& \frac{1}{(2 \pi )^{3/2}} \int d^3 r \, {\bf B} ({\bf r}, 0) \, e^{i {\bf k} \cdot {\bf r} } \label{b0}
\end{eqnarray}
These transforms are then inverted multiplied by terms as $e^{- i ( {\bf k} \cdot {\bf r} - \omega \, t )}$ and $e^{-i ( {\bf k} \cdot {\bf r} + \omega \, t )}$, where $\omega = c \, k$, so that the time evolution of the fields is obtained. In our cases we get the exact solutions,
\begin{eqnarray}
{\bf B} ({\bf r}, t) &=& \frac{\sqrt{a}}{\pi L_{0}^2} \, \frac{Q \, {\bf H}_{1} + P \, {\bf H}_{2}}{(A^2 + T^2 )^3} \label{knot10} \\
{\bf E} ({\bf r}, t) &=& \frac{\sqrt{a} c}{\pi L_{0}^2} \, \frac{Q \, {\bf H}_{4} - P \, {\bf H}_{3}}{(A^2 + T^2 )^3} \label{knot11}
\end{eqnarray}
where the quantities $A$, $P$, $Q$ are defined by
\begin{eqnarray}
A &=& \frac{R^{2}-T^{2}+1}{2}, \label{knot111} \\ 
P &=& T(T^{2}-3A^{2}), \label{knot112} \\ 
Q &=& A(A^{2}-3T^{2}), \label{knot113}
\end{eqnarray}
and the vectors ${\bf H}_{1}$, ${\bf H}_{2}$, ${\bf H}_{3}$ and ${\bf H}_{4}$ are
\begin{eqnarray}
{\bf H}_{1} &=& \left( -n \, XZ + m \, Y + s \, T \right) \, {\bf u}_{x} + \left( -n \, YZ -m \, X -l \, TZ \right) \, {\bf u}_{y} \nonumber \\
&+& \left( n \, \frac{-1 - Z^2 + X^2 + Y^2 +T^2}{2} + l \, TY \right) \, {\bf u}_{z} \nonumber \\
{\bf H}_{2} &=& \left( s \, \frac{1+X^2 -Y^2-Z^2-T^2}{2} -m \, TY \right) \, {\bf u}_{x} + \left( s \, XY - l \, Z + m \, TX \right) \, {\bf u}_{y} \nonumber \\
&+& \left( s \, XZ + l \, Y + n \, T \right) \, {\bf u}_{z} \nonumber \\
{\bf H}_{3} &=& \left( -m \, XZ + n \, Y + l \, T \right) \, {\bf u}_{x} + \left( -m \, YZ -n \, X -s \, TZ \right) \, {\bf u}_{y} \nonumber \\
&+& \left( m \, \frac{-1 - Z^2 + X^2 + Y^2 +T^2}{2} + s \, TY \right) \, {\bf u}_{z} \nonumber \\
{\bf H}_{4} &=& \left( l \, \frac{1+X^2 -Y^2-Z^2-T^2}{2} - n \, TY \right) \, {\bf u}_{x} + \left( l \, XY - s \, Z + n \, TX \right) \, {\bf u}_{y} \nonumber \\ 
&+& \left( l \, XZ + s \, Y + m \, T \right) \, {\bf u}_{z}. \label{knot12}
\end{eqnarray}
One important feature of these solutions is the finite value of the electromagnetic energy. The energy is related to the integer numbers
$n$, $m$, $l$, $s$, that characterize the complex scalar fields $\phi$ and $\theta$ by
\begin{equation}
{\cal E} = \int d^3 r \, \left( \frac{\varepsilon_{0} \, E^2}{2} + \frac{B^2}{2 \mu_{0}} \right) = \frac{a}{2 \mu_{0} L_{0}} (n^2 + m^2 + l^2 + s^2) . \label{knot13}
\end{equation}
The linear momentum of these knotted solutions can also be obtained from the Poynting vector ${\bf E} \times {\bf B} / \mu_{0}$ and results
\begin{equation}
{\bf p} = \int \varepsilon_{0} \, {\bf E} \times {\bf B} = \frac{a}{2 c \mu_{0} L_{0}} \, (l n + m s) \, {\bf u}_{y} . \label{knot14}
\end{equation}
A way to picture the evolution of these fields is to plot Poynting vectors at some points in space at different times. In figure \ref{poynting}, we see the behaviour of these electromagnetic fields for instants of time $T=-2, -1, 0, 1, 2$ (where $T= c t /L_{0}$). There is a focusing effect of the energy density flux at $T=0$. In the figure we have plotted the case $(n,m) = (3,4)$ for the initial magnetic field lines, and $(l,s) = (2,3)$ for the initial electric field lines. The magnetic and electric fields are given by equations (\ref{knot10}) and (\ref{knot11}) with $n=3$, $m=4$, $l=2$, $s=3$.

With expressions (\ref{knot10}) and (\ref{knot11}), one can also calculate the angular momentum of these solutions, that reads
\begin{equation}
{\bf J} = \int \varepsilon_{0} \, {\bf r} \times \left(  {\bf E} \times {\bf B} \right) = \frac{a}{2 c \mu_{0}} \, 
(l m + n s) \, {\bf u}_{y} , \label{knot15} 
\end{equation}
so the electromagnetic energy flux goes along the $y$-axis, and its angular momentum is parallel to its linear momentum.

Other interesting dynamical quantities are related to the Lorentz invariants associated to any electromagnetic field in vacuum. These
invariants are ${\bf E} \cdot {\bf B}$ and $E^2 - c^2 \, B^2$, that can also be obtained from the square of the Riemann-Silberstein 
complex vector ${\bf F} = {\bf E} + i \, c \, {\bf B}$, that is a very useful tool to compute many dynamical aspects of the electromagnetic fields in vacuum \cite{Birula,Besieris}. From equations (\ref{knot10})-(\ref{knot11}),
\begin{equation}
{\bf F} ({\bf r}, t) = \frac{\sqrt{a} c}{\pi L_{0}^2} \, \frac{Q \, \left( {\bf H}_{4} + i \, {\bf H}_{1} \right) - P \, 
\left( {\bf H}_{3} - i \, {\bf H}_{2} \right)}{(A^2 + T^2 )^3} .
\label{knot16}
\end{equation}
For this class of knotted electromagnetic fields, with $A=( R^{2}-T^{2}+1)/2$,
\begin{eqnarray}
{\bf E} \cdot {\bf B} &=& \frac{ac}{\pi^2 L_{0}^4} \, \frac{1}{(A^2 + T^2 )^4} \, ( (m s - n l ) (A^2 + T^2) XZ \nonumber \\ 
&+& (ml-ns) (A^2 +T^2) (1-A) Y + 2 (l s - m n)  A T (T^2 - A^2) )     \label{knot17} \\
E^2 - c^2 \, B^2 &=& \frac{ac^2}{\pi^2 L_{0}^4} \, \frac{1}{(A^2 + T^2 )^4} \, ( (n^2 - m^2 ) (A^2 + T^2 ) (X^2 + Y^2 ) \nonumber \\ &+& (s^2 - l^2 ) (A^2 + T^2 ) (Y^2 + Z^2 ) + 4 (m^2 - s^2) A^2 T^2 - (n^2 - l^2 ) (A^2 - T^2 )^2 ) \label{knot18}
\end{eqnarray}

\section{Special Properties of the torus-knotted fields}
The choice of the positive integer numbers $n, m, l, s$ determines some properties of the knotted electromagnetic fields. They are related to the invariants of the electromagnetic fields and the magnetic and electric helicities. In this section we will state them.  

The class of knotted electromagnetic fields has the following properties:
\begin{itemize}
\item{\it Property C1}. If the positive integer numbers $n, m, l, s$ are equal, i. e. satisfy the condition $n=m=l=s$, then ${\bf E} \cdot {\bf B} = 0$ and $E^2 - c^2 \, B^2 = 0$. In terms of the Riemann-Silberstein vector ${\bf F}$ defined in (\ref{knot16}), ${\bf F} \cdot {\bf F} =0$.
\item{\it Property C1}$^{\, \prime}$. If the condition $n=m=l=s$ is not fulfilled, then ${\bf E} \cdot {\bf B} \neq 0$ and $E^2 - c^2 \, B^2 \neq 0$. 
\end{itemize} 
Remarkably, and with the help of equation (\ref{knot17}), the time behaviour of the helicities of these electromagnetic fields can be exactly
computed. The result, in terms of the dimensionless parameter $T = c t / L_{0}$, is
\begin{eqnarray}
h_{m} (t) &=& \frac{1}{2 \mu_{0}} \int d^3 r \, {\bf A} \cdot {\bf B} = \frac{a}{4 \mu_{0}} \, \left[ (mn + ls) + (mn - ls) \, \frac{1-6 T^2 + T^4}{(1+T^2)^4} \right] , \label{knot20} \\ 
h_{e} (t) &=& \frac{\varepsilon_{0}}{2} \int d^3 r \, {\bf C} \cdot {\bf E} = \frac{a}{4 \mu_{0}} \, \left[ (mn + ls) - (mn - ls) \, \frac{1-6 T^2 + T^4}{(1+T^2)^4} \right] \label{knot21} .
\end{eqnarray}
So for this class of knotted electromagnetic fields we have
\begin{itemize}
\item{\it Property C2}. If, at $t=0$, we have $h_{m}=h_{e}$, then for every time we have $h_{m} = h_{e}$ and both helicities are constant. That means that $ m n = l s$, thus
\begin{equation}
\int d^3 r \, \operatorname{Im}({\bf F} \cdot {\bf F})= 2 c \, \int d^3 r \, {\bf E} \cdot {\bf B} = 0.
\label{knot19}
\end{equation}
As a consequence, the magnetic and electric helicities will be constant during the time evolution of the fields.
\item{\it Property C2}$^{\, \prime}$. If, at $t=0$, we have $h_{m} \neq h_{e}$, then both helicities are not constant but, for large $T$, both reach the same value (due to a mechanism called exchange of helicities in \cite{Arr11}).
\end{itemize}

\section{Study of the field lines of knotted electromagnetic fields with constant helicities ($l=n$ and $s=m$)}
In the previous section we have seen how to construct analytically
different electromagnetic fields in vaccum for which, at $t=0$, all
the magnetic lines and all the electric lines are linked torus knots
in the space $R^3$, except a zero-measure set. For this fields, at $t=0$ all the magnetic lines
are the same torus knot, and all the electric lines are the same torus
knot. But in general the kind of torus knot corresponding to the
magnetic lines may be different from the kind of torus knot
corresponding to the electric lines.

Next we are going to restrict ourselves to the case in which, at
$t=0$, both the magnetic and the electric lines are the same kind of
torus knot. This means that, in equations
(\ref{knot10})-(\ref{knot12}), we take $l=n$ and $s=m$. Consequently,
the fields satisfy Properties {\it C1}$^{\, \prime}$ and {\it C2}, so
that the magnetic field is not orthogonal to the electric field but
the magnetic and the electric helicities are equal and constant in
time. In reference \cite{Irv08}, a different method to construct these
kind of fields was proposed, although that method does not provide the 
more general solutions (\ref{knot10})-(\ref{knot12}) presented here in the
previous sections.

\begin{figure}
\centering
\includegraphics[width=0.3\textwidth]{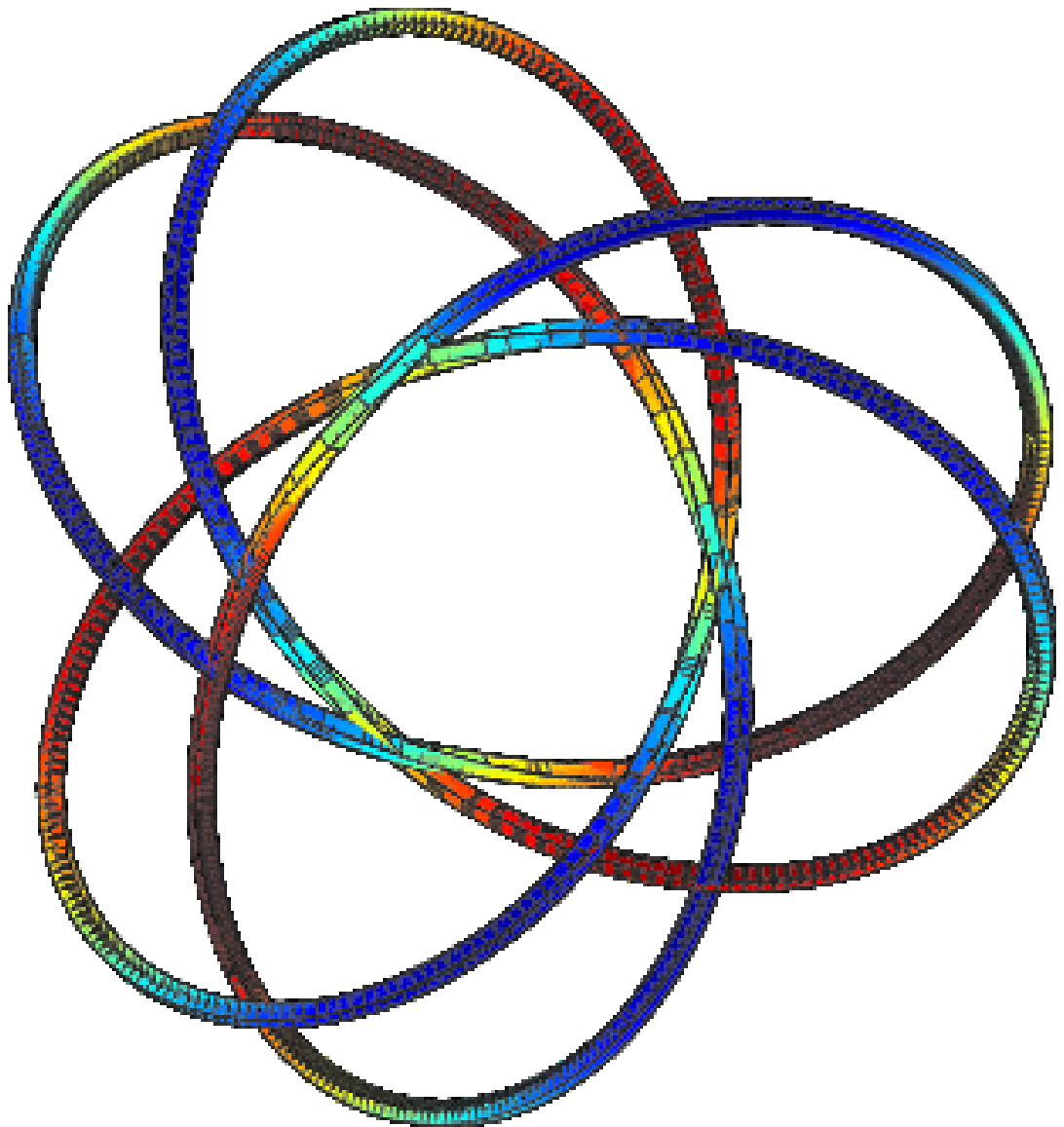}
\includegraphics[width=0.3\textwidth]{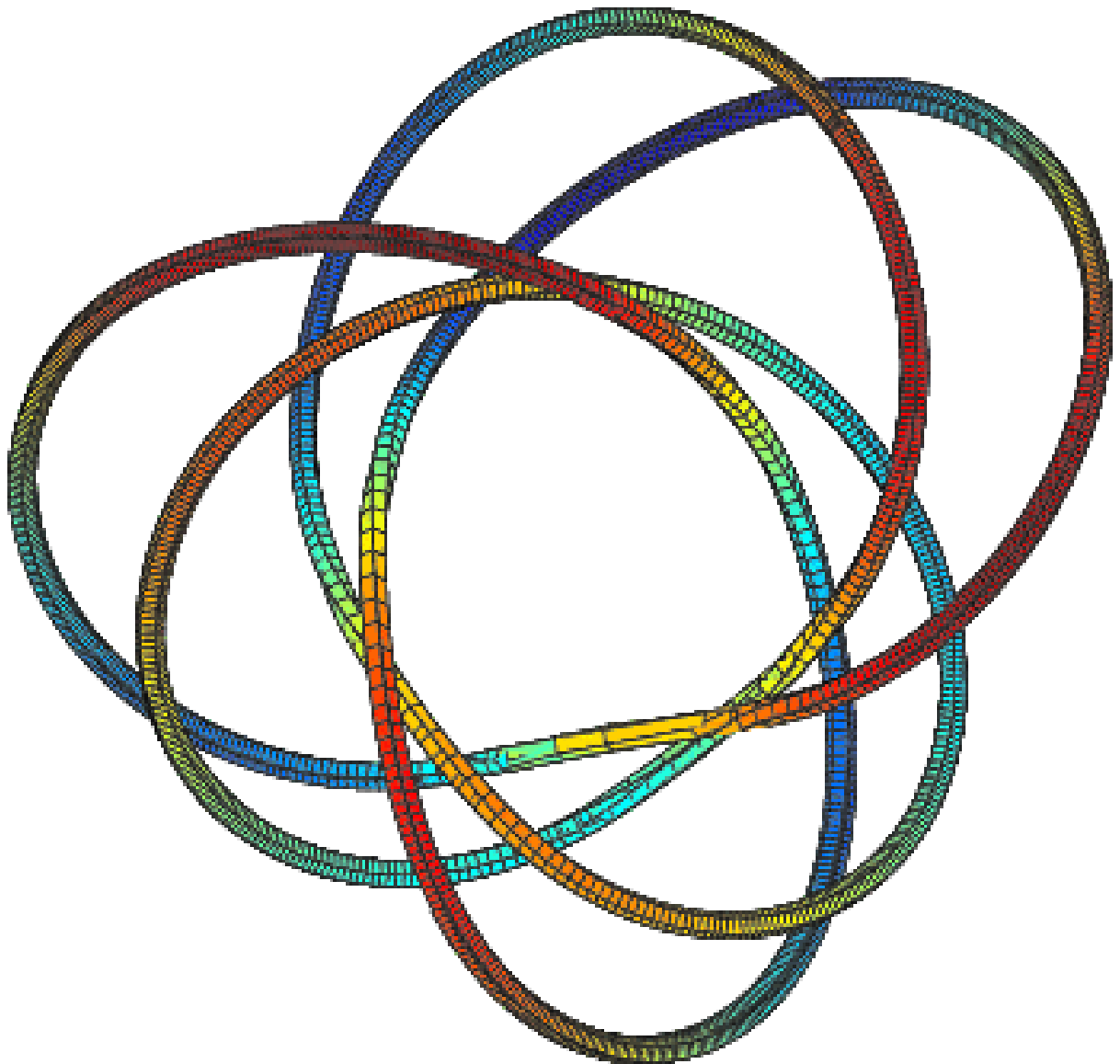}
\includegraphics[width=0.3\textwidth]{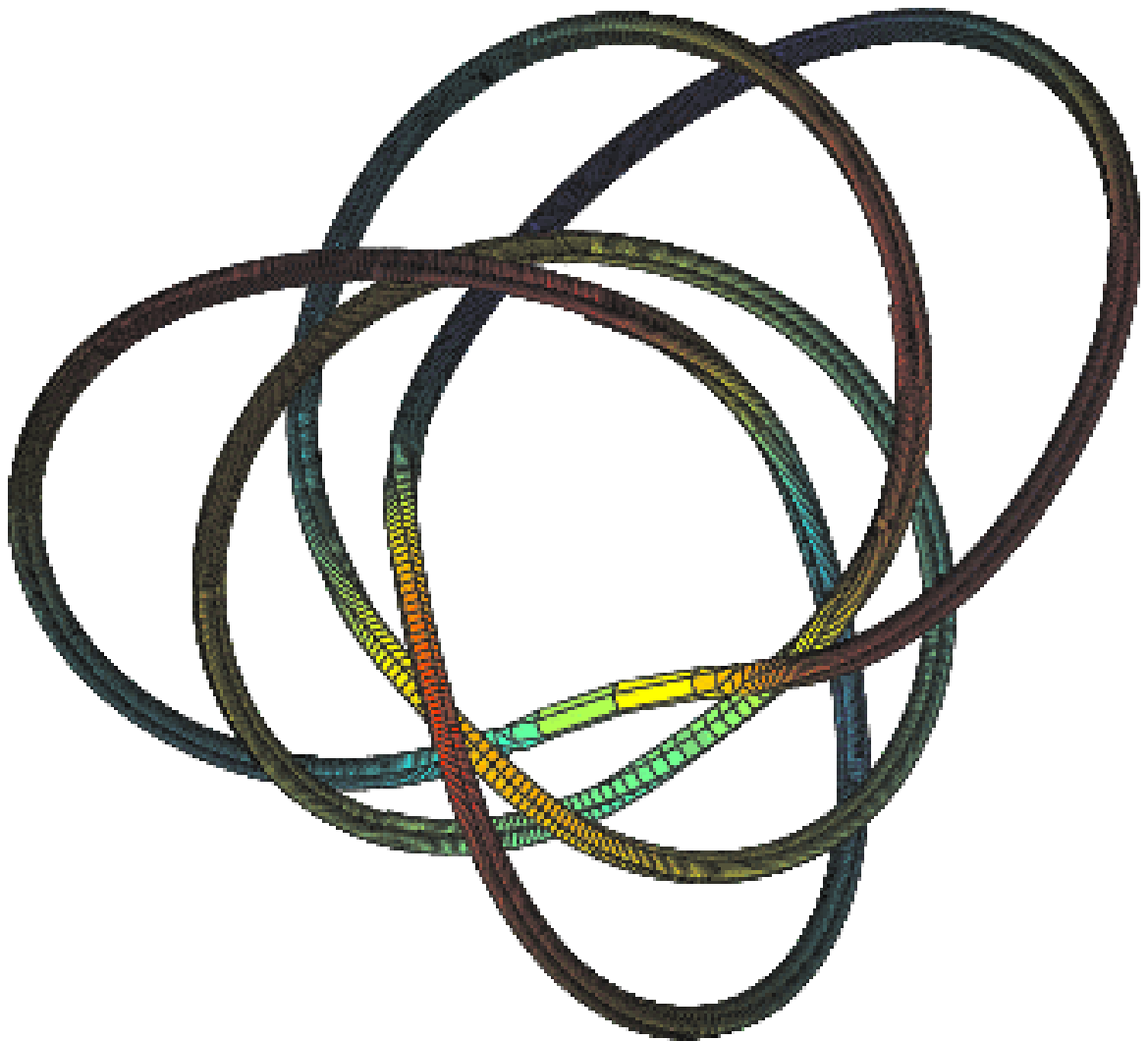}
\includegraphics[width=0.3\textwidth]{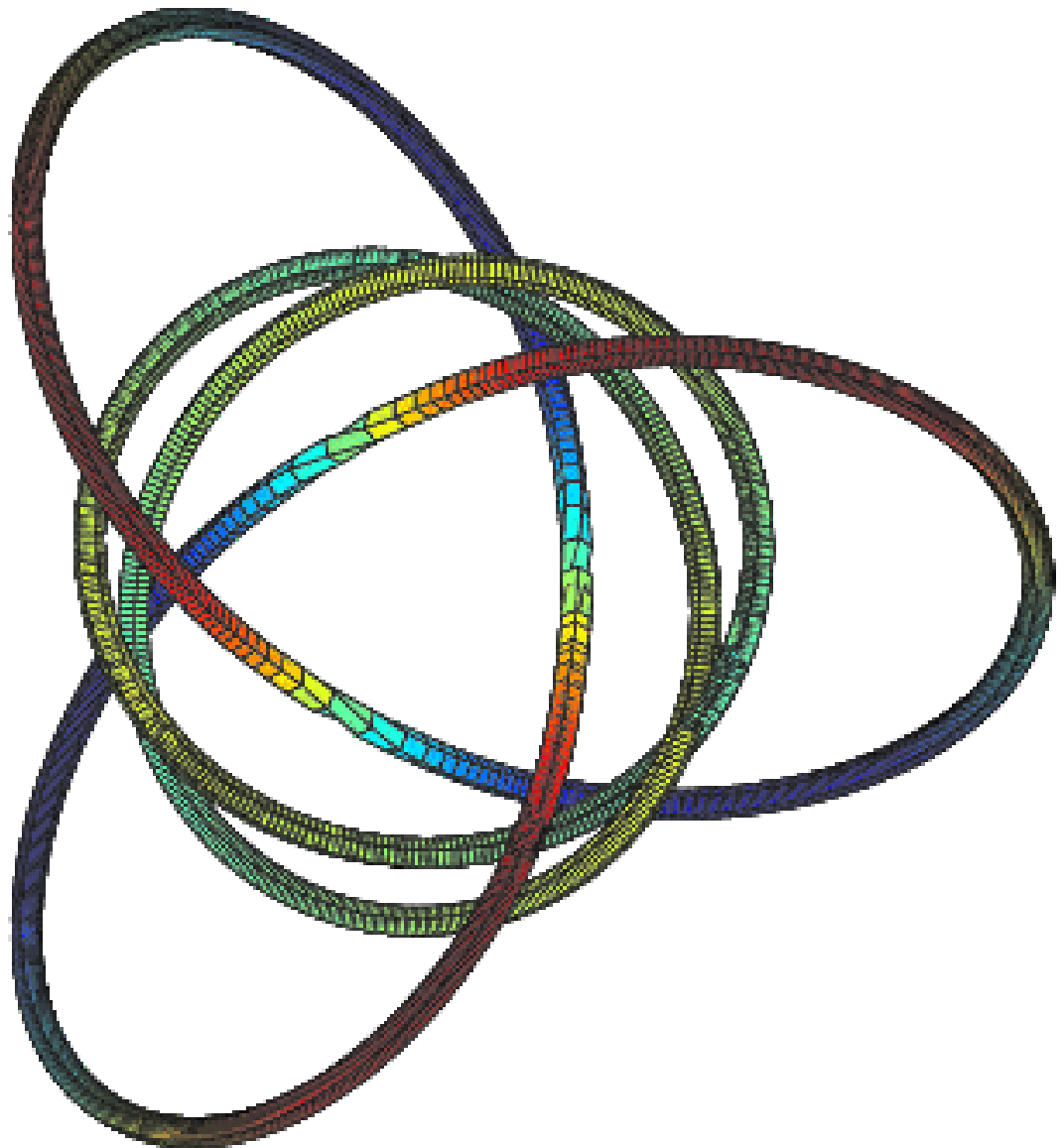}
\includegraphics[width=0.3\textwidth]{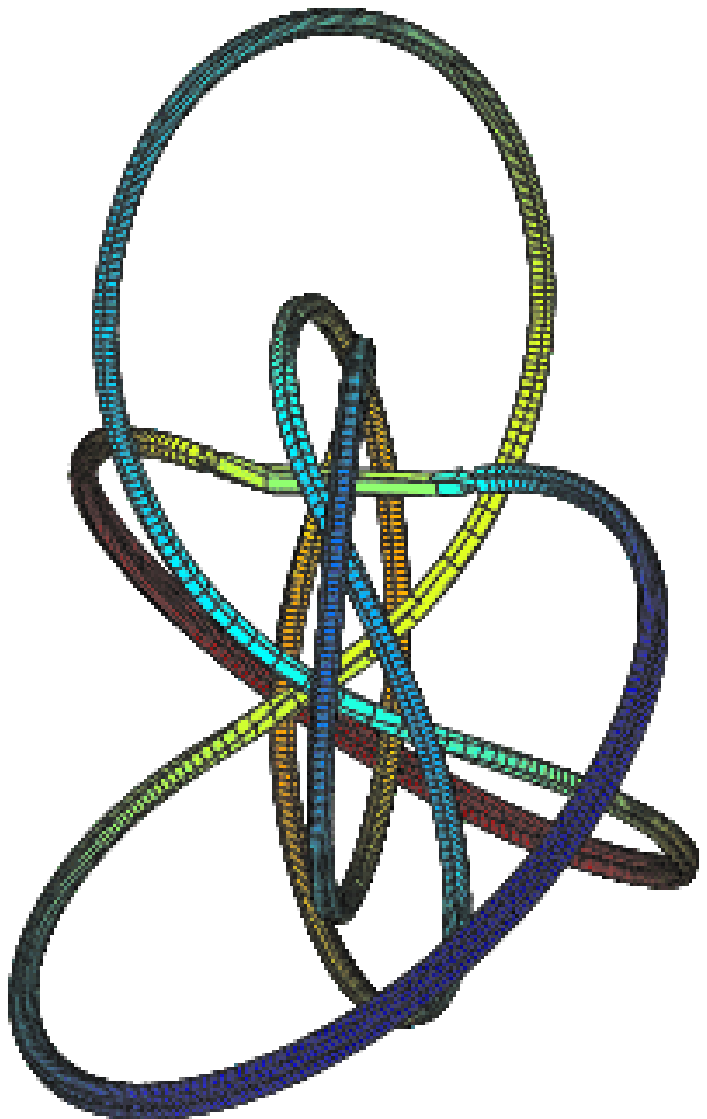}
\includegraphics[width=0.3\textwidth]{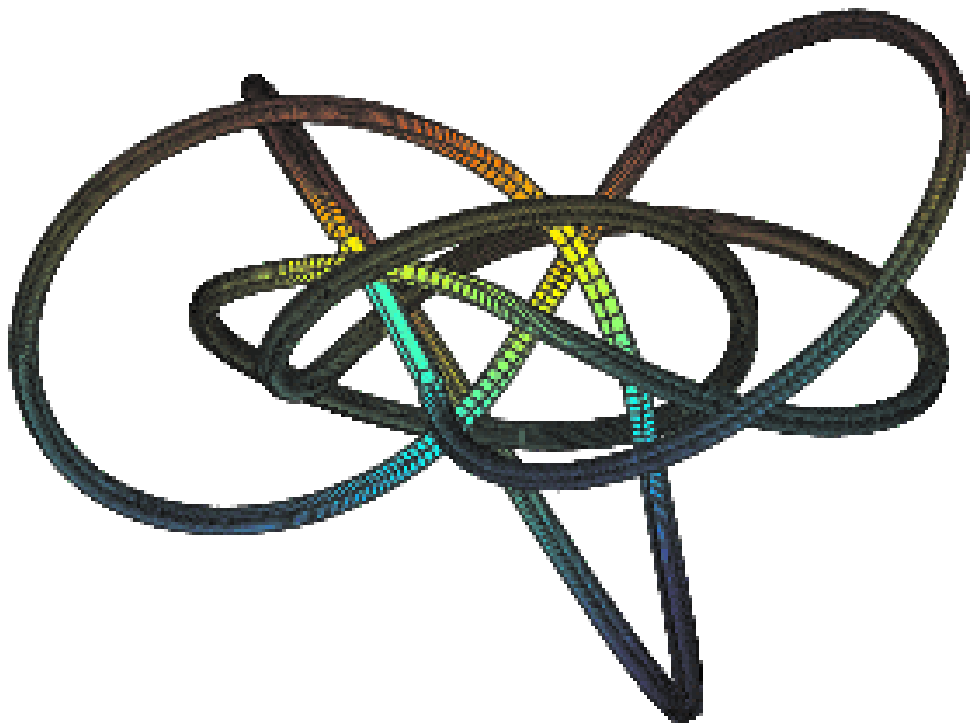}
\caption{Time evolution of field lines for the (3,2) torus knot or trefoil. We plot two magnetic (up) 
and electric (down) lines for $T=0$, $T=0.015$ and $T=0.02$ (left to right). Lines are obtained numerically 
by integrating the explicit time dependent expressions (\ref{knot10})-(\ref{knot11}) of the magnetic and the electric field. 
Numerical integration shows that at $T=0$ all the magnetic lines (and the electric lines) are trefoils 
except 2 lines (as always occurs in torus knots): the straight line $X=Y=0$ and the circle $X^2+Y^2=1$. 
Moreover, the trefoils are linked to each other so that the linking number is $n m =6$. When $T>0$ 
we find numerically magnetic (and electric) lines that are linked trefoils.} 
\label{figtrefoil}
\end{figure}

\begin{figure}
\centering
\includegraphics[width=0.3\textwidth]{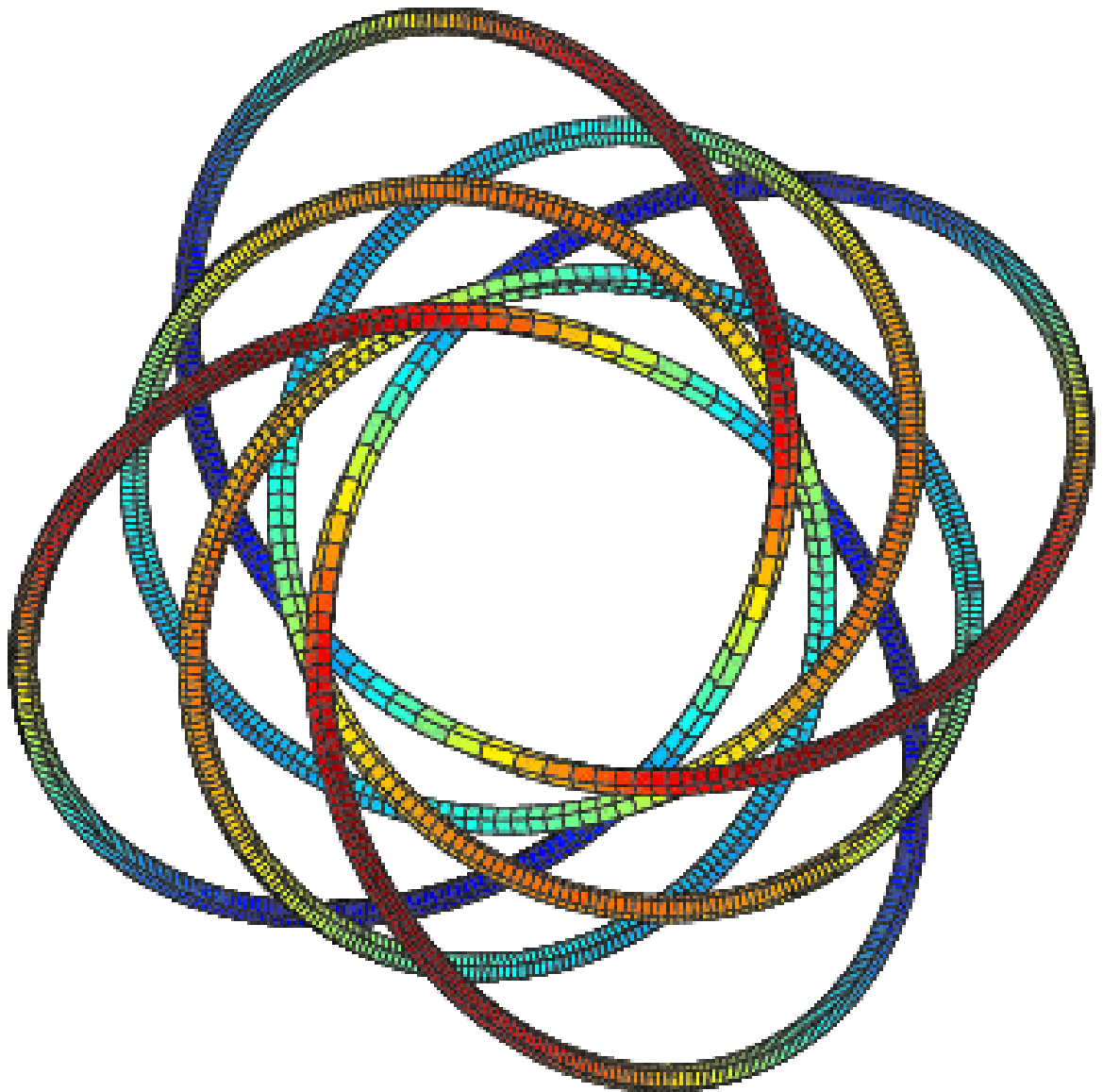}
\includegraphics[width=0.3\textwidth]{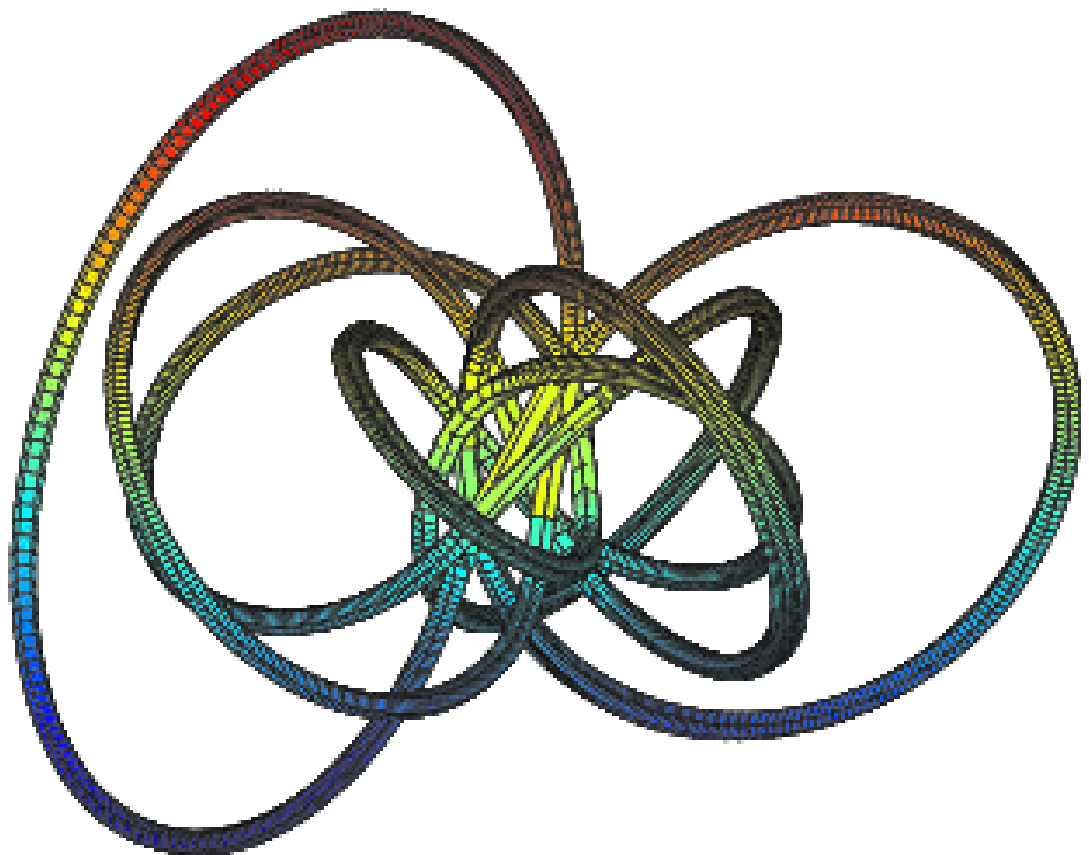}
\includegraphics[width=0.3\textwidth]{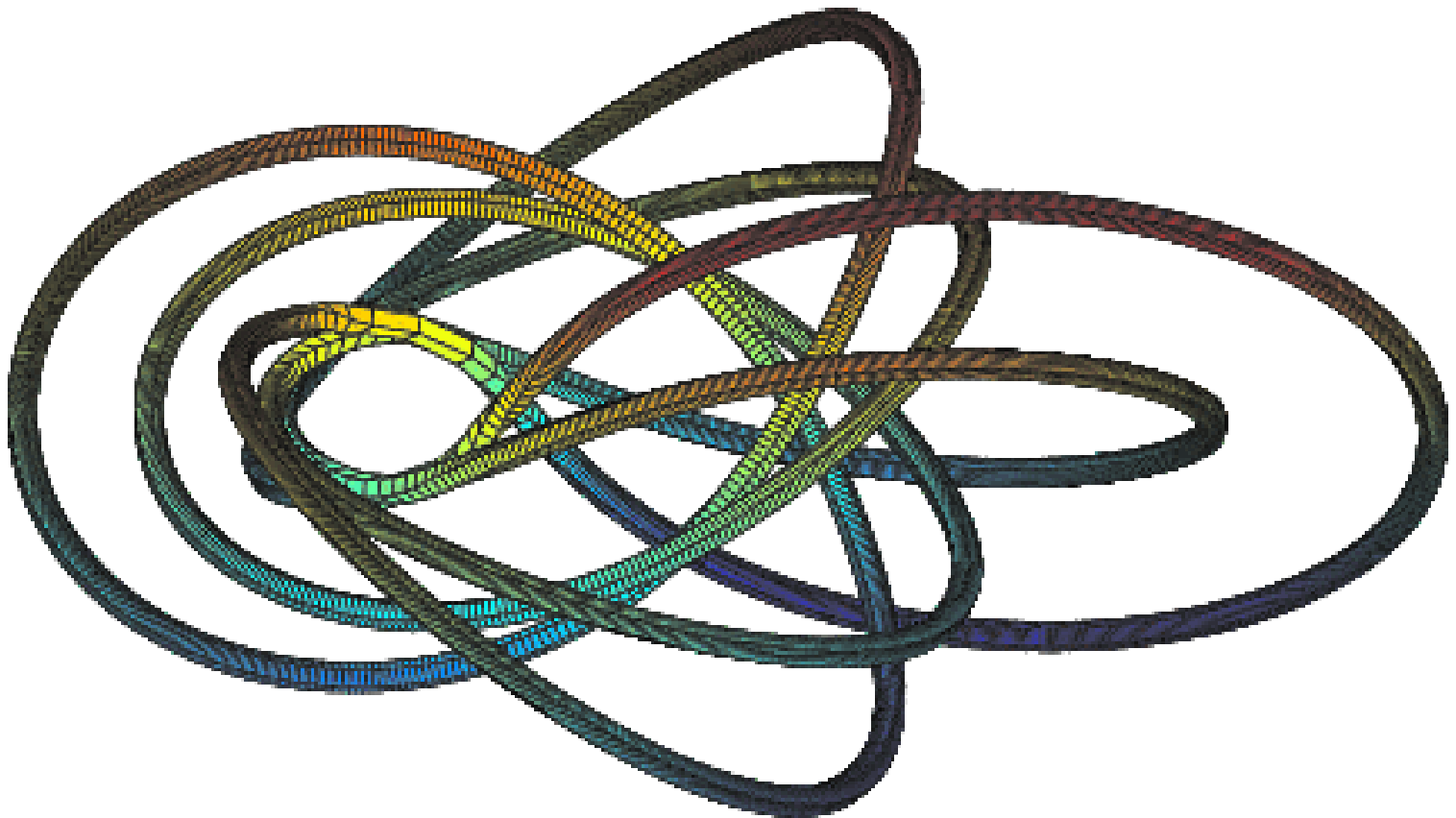}
\includegraphics[width=0.3\textwidth]{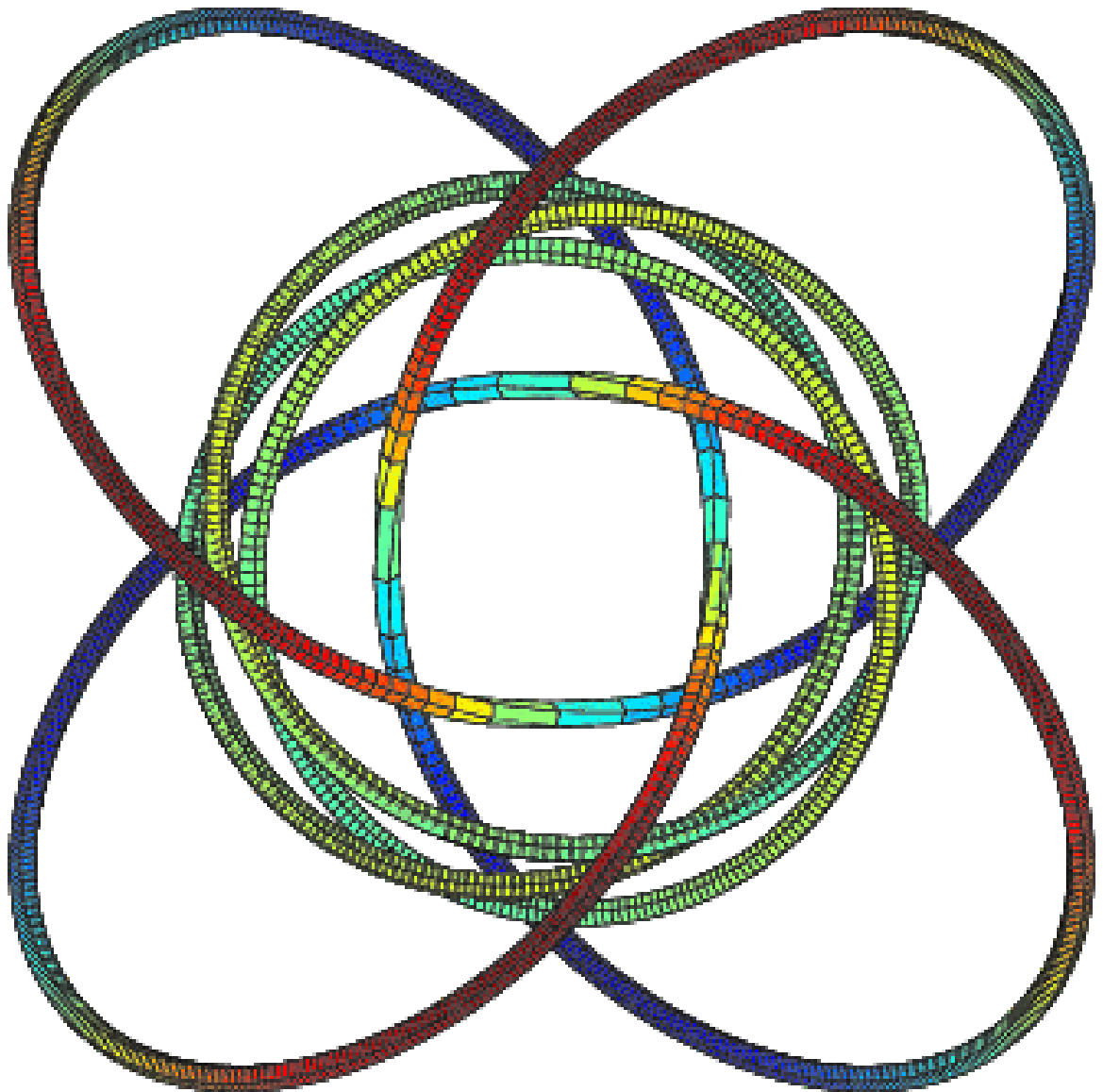}
\includegraphics[width=0.3\textwidth]{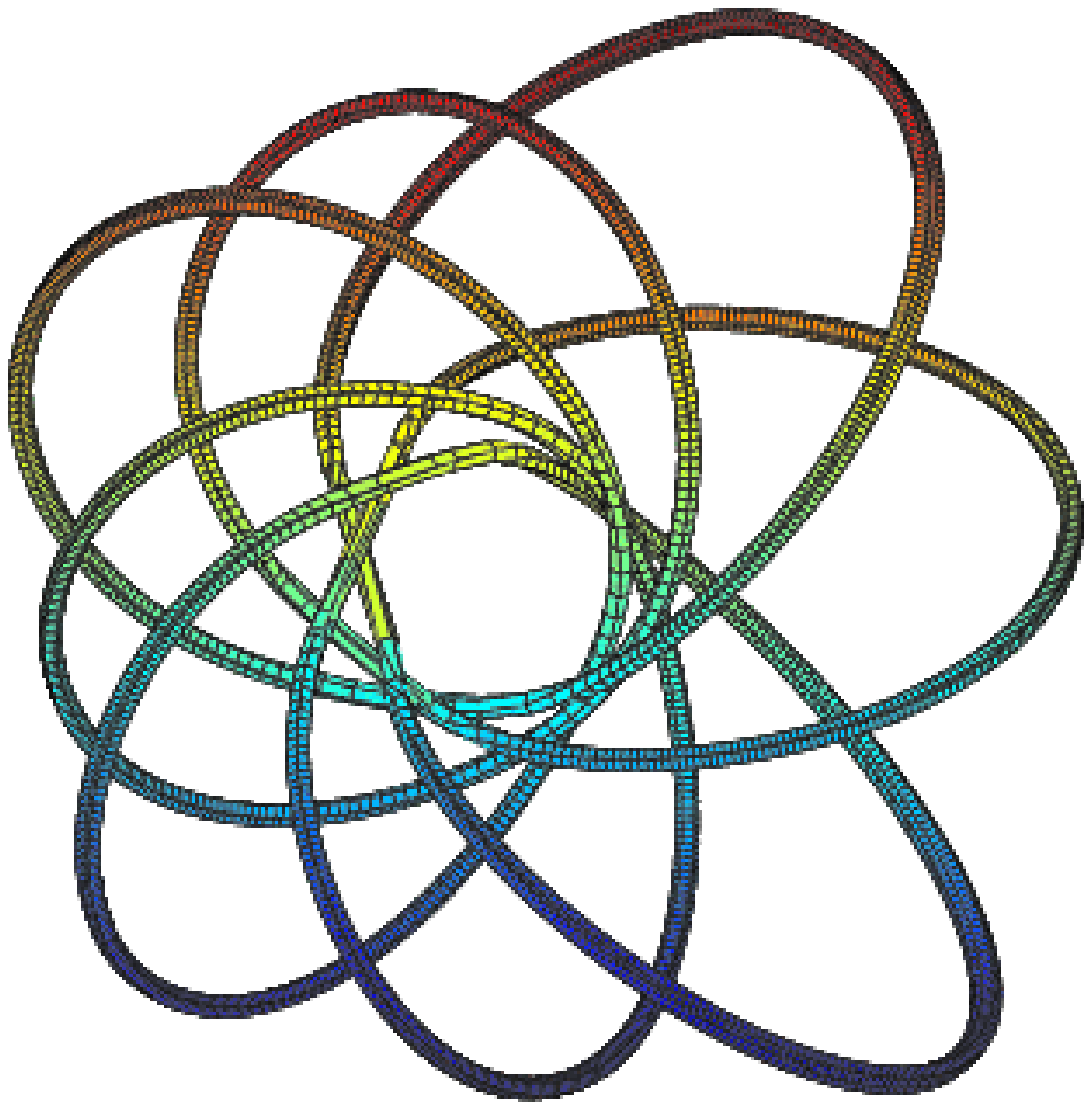}
\includegraphics[width=0.3\textwidth]{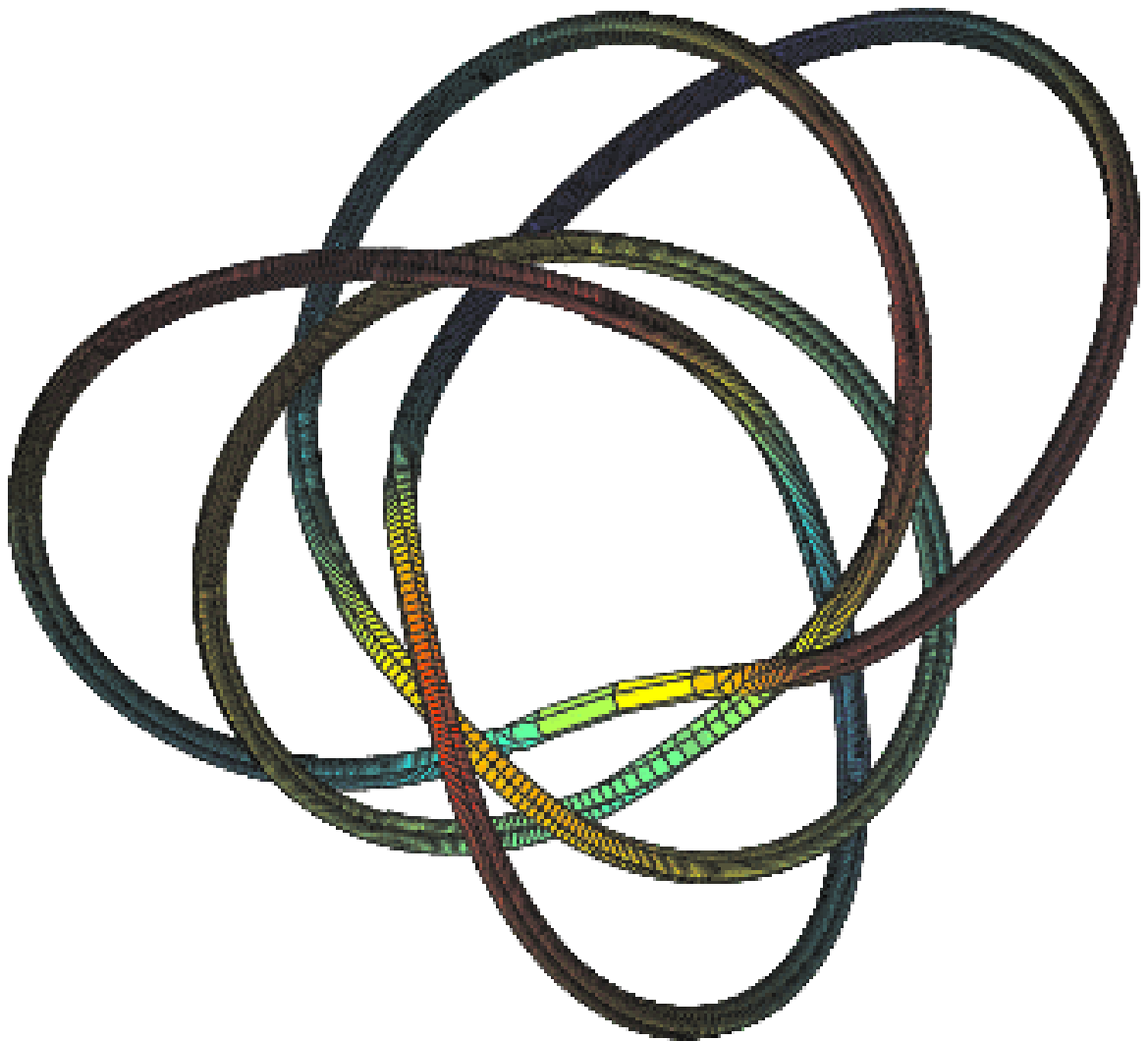}
\caption{Time evolution of field lines for the (4,3) torus knot. We plot two magnetic (up) and electric 
(down) lines for $T=0$, $T=0.015$ and $T=0.02$ (left to right). Lines are obtained numerically by integrating 
the explicit time dependent expressions (\ref{knot10})-(\ref{knot11}) of the magnetic and the electric field. At $T=0$ all the field
lines are linked $(4,3)$ torus knots and the linking number is $n m =12$. When $T>0$ we find $(4,3)$ 
linked torus knots.} 
\label{figluck}
\end{figure}

\begin{figure}
\centering
\includegraphics[width=0.3\textwidth]{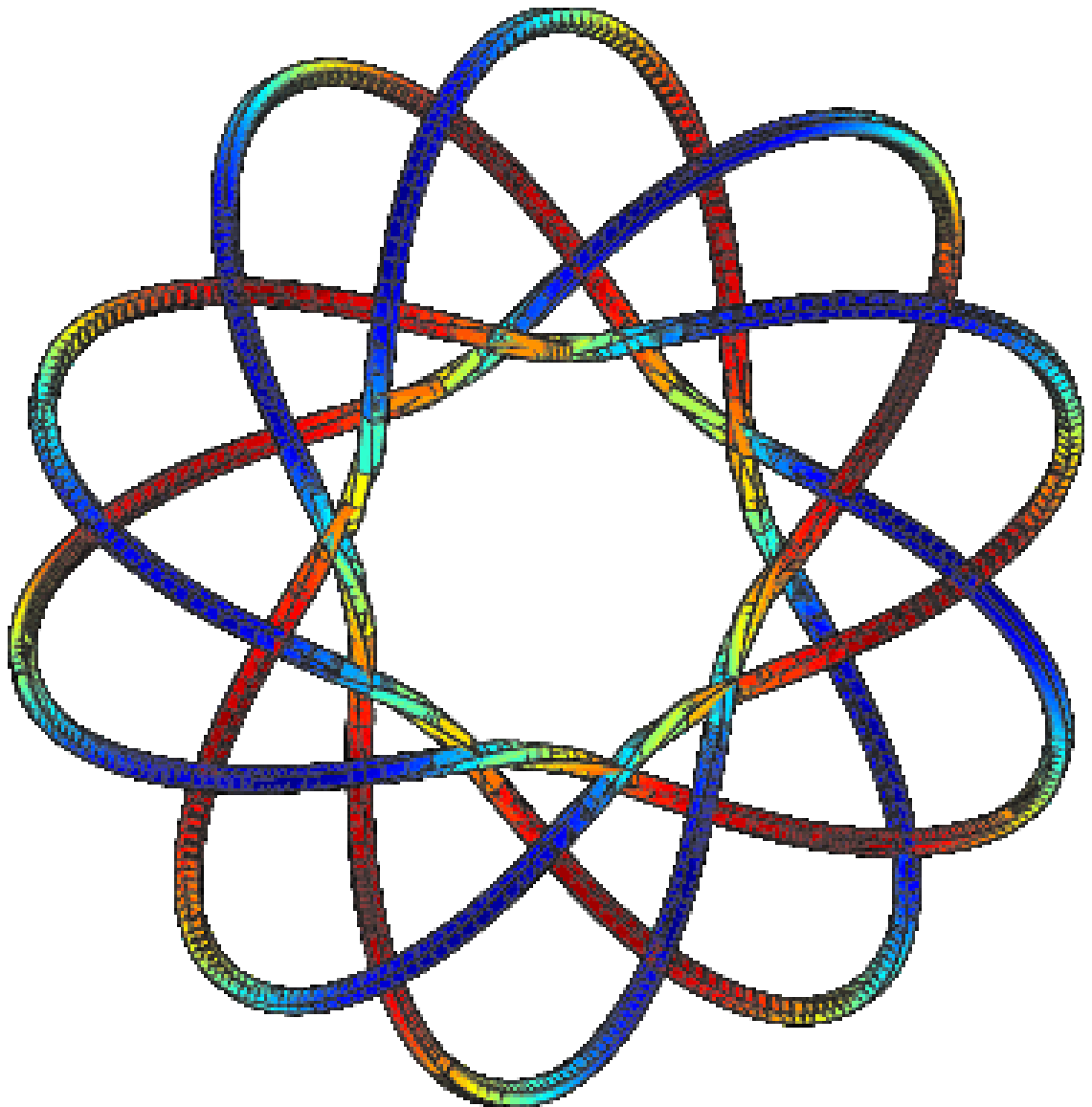}
\includegraphics[width=0.3\textwidth]{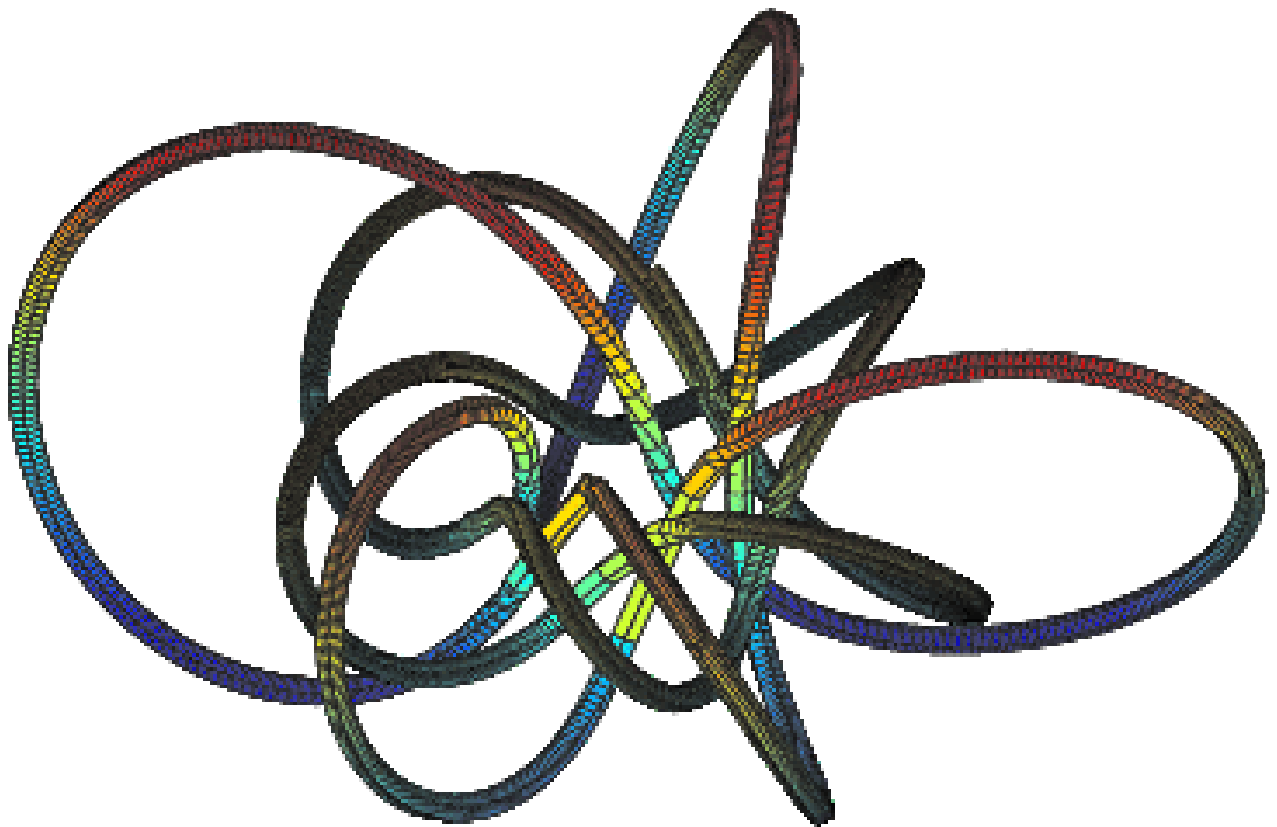}
\includegraphics[width=0.3\textwidth]{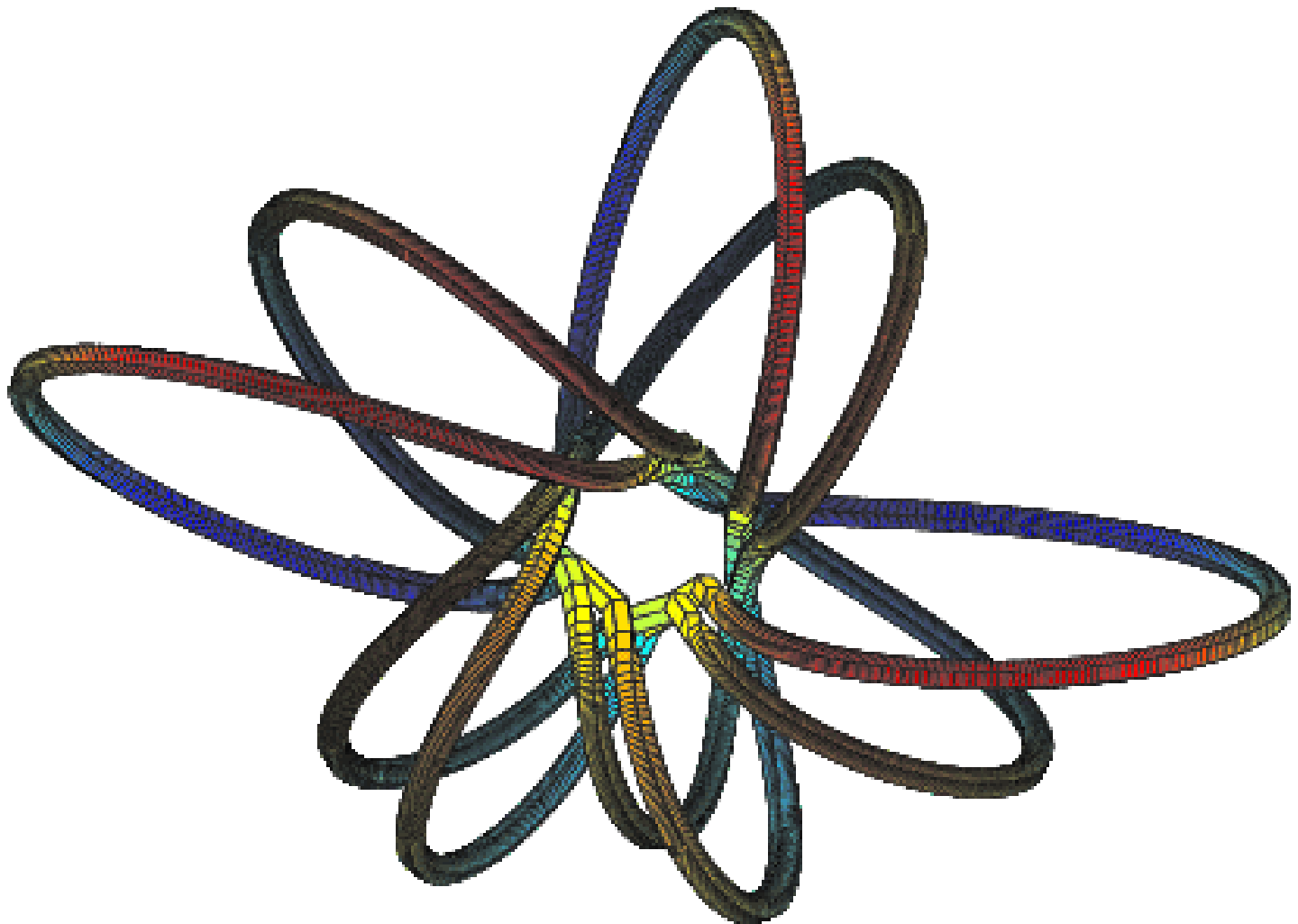}
\includegraphics[width=0.3\textwidth]{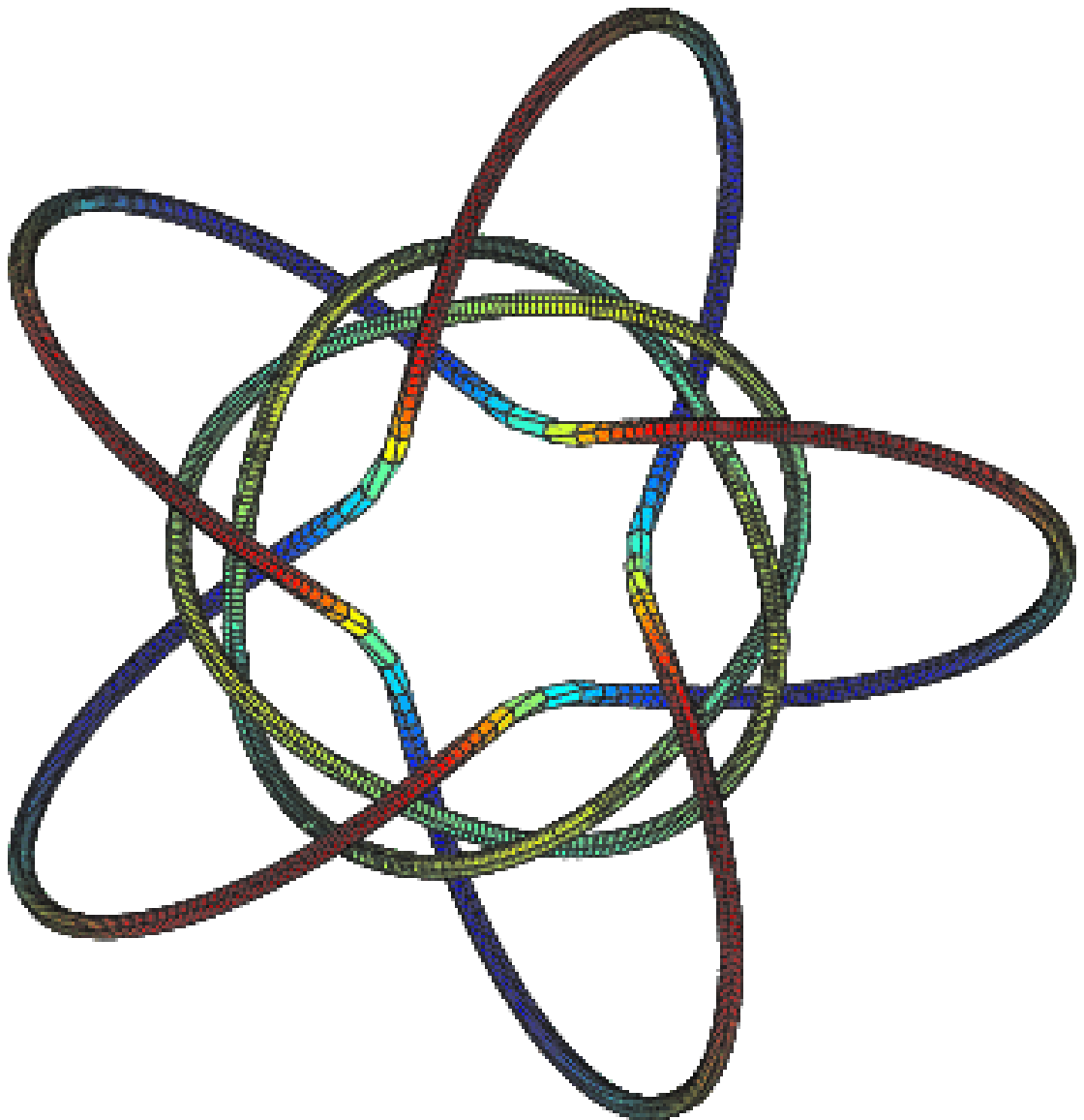}
\includegraphics[width=0.3\textwidth]{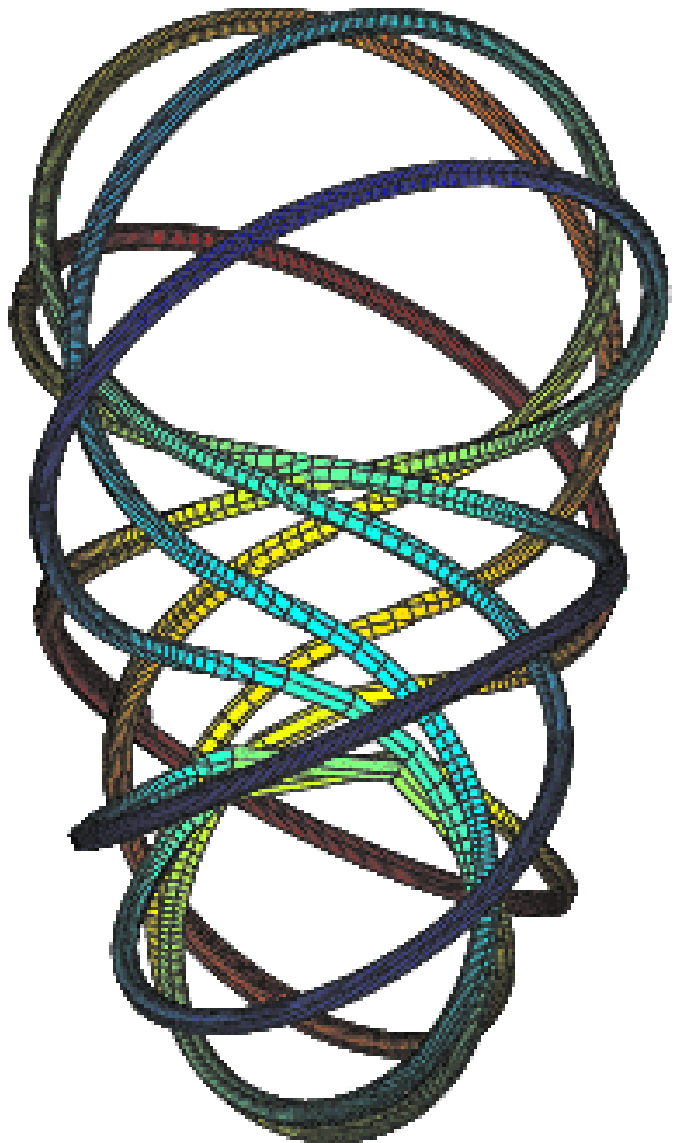}
\includegraphics[width=0.3\textwidth]{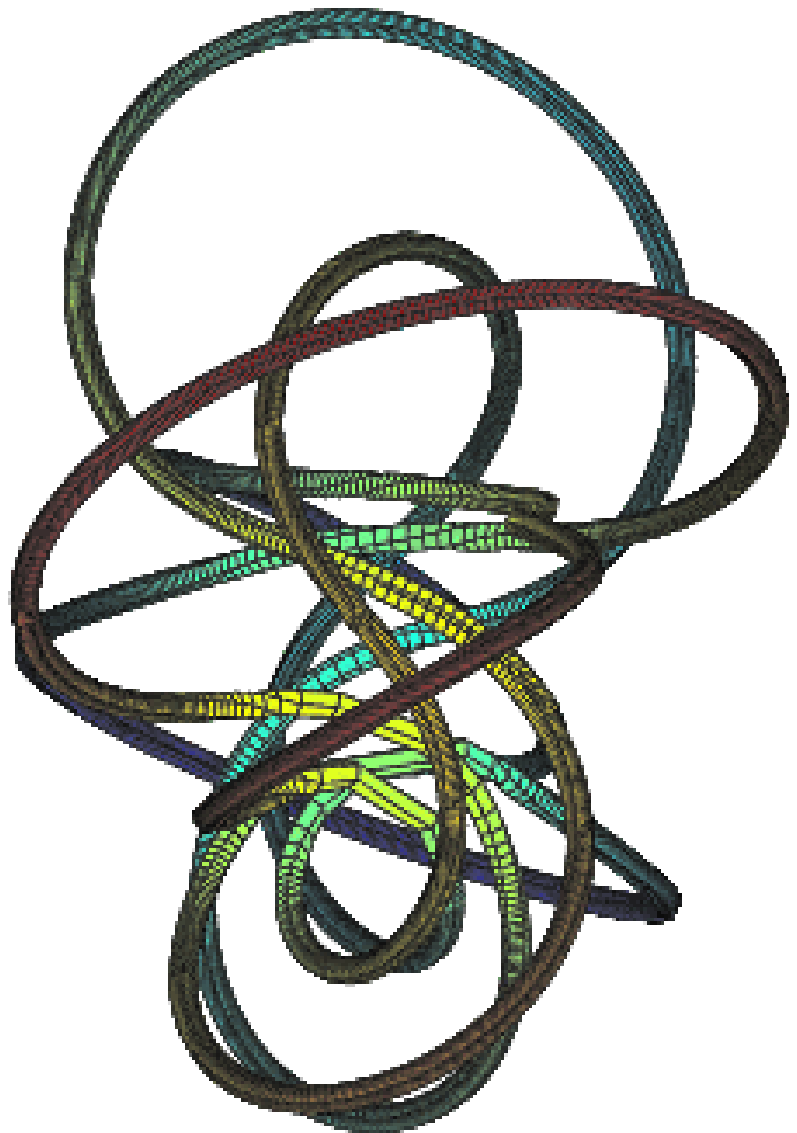}
\caption{Time evolution of field lines for the (5,2) torus knot. We plot two magnetic (up) and electric (down) 
lines for $T=0$, $T=0.015$ and $T=0.02$ (left to right). Lines are obtained numerically by integrating the 
explicit time dependent expressions (\ref{knot10})-(\ref{knot11}) of the fields. At $T=0$ all the field lines are linked $(5,2)$ 
torus knots and the linking number is $n m =10$. When $T>0$ we find $(5,2)$ linked torus knots.} 
\label{figstar}
\end{figure}

To see how these solutions give different torus knots, we plot in
figure \ref{figtrefoil} some magnetic and electric lines for the
particular case $n=3$, $m=2$. At $T=0$ the lines are linked trefoils
and the linking number is $H(\phi)=H(\theta)= n m =6$. When $T>0$
we also find a set of linked trefoils, although not all the lines
are linked trefoils at $T > 0$. In figure \ref{figluck} we see the
behaviour of magnetic and electric lines in the case $n=4$, $m=3$. At
$T=0$, all the lines are linked $(4,3)$ torus knots and the linking
number is $n m =12$. For $T>0$ we find lines with the same
topology. The same is done in figure \ref{figstar} with $n=5$, $m=2$
and linking number $10$. In all these figures it is clear
that plotted curves are linked knots. When $T=0$, these are the only
kind of curves that can be found since these lines are level curves of
the scalar fields (\ref{knot3})-(\ref{knot4}). As $T$ increases, numerical 
evidence prevent us to rule out the existence of some open field lines 
together with the knotted closed ones. A similar behaviour was
found also numerically in \cite{Irv08}.

\section{The special case $n=m=l=s$: Hopf-Ra\~nada electromagnetic knot}
In the special case in which we set solutions of the equations 
(\ref{knot11})-(\ref{knot12}) for which $n=m=l=s$, we have a situation in 
which Properties {\it C1} and {\it C2} hold. These particular solutions, called 
Hopf-Ra\~nada electromagnetic knots \cite{Trautman,Ran89,Ran95,Ran97,Arr10,Irv08}, 
are based on the Hopf fibration. They constitute a kind of basis for general 
electromagnetic fields in vacuum with nontrivial topology of the field lines.

In this case, the linkage of magnetic or electric lines is conserved
in time. It has been analytically proof that all the magnetic and
electric lines remain closed and linked for any value of $T$ since
there is an explicit expression for the scalar fields $\phi$ and
$\theta$ of equations (\ref{knot1})-(\ref{knot2}) for any time. These
time dependent expressions were firstly found in \cite{Ran97} and they
are
\begin{eqnarray}
\phi &=& \frac{(A \, X - T \, Z ) + \, i ( A \, Y + T \, (A-1) )}{(A \, Z + T \, X ) + \, i ( A \, (A-1) - T \, Y)} , \label{knot22} \\ 
\theta &=& \frac{(A \, Y + T \, (A-1) ) + \, i ( A \, Z + T \, X )}{(A \, X - T \, Z ) + \, i ( A \, (A-1) - T \, Y)} , \label{knot23}
\end{eqnarray}
where $A$ is given in equation (\ref{knot111}). The time evolution of both scalars
is smooth and this implies that the field lines remain closed and the topology is
conserved.

In the general case with different values of the integer numbers $n$, $m$, $l$, $s$,
it is not possible to find the scalar fields $\phi$ and $\theta$ for $T > 0$. Due
to this fact, we have found numerically open curves for $T > 0$.

\section{Conclusions}
In conclusion, we have presented new exact solutions of Maxwell's equations in vacuum such that, at 
a given initial time, satisfy that all the magnetic lines and all the electric lines are torus knots. 
Expressions (\ref{knot10})-(\ref{knot12}) generalize previous works on knotted electromagnetic
fields. We have found some properties satisfied by these solutions, as the values of the Lorentz
invariants and the magnetic and electric helicities. The Lorentz invariants associated to these
solutions depend on the position and time, and the magnetic and electric helicities depend on time.
However, we have shown that, when time is large, the magnetic and electric helicities become 
equal.

We have computed numerically the field lines corresponding to cases in which the magnetic and electric
helicities are constant in time. These are situations in which the initial electric lines and the
initial magnetic lines are the same kind of torus knots. When time evolves, we
have observed in the numerical computations of the field lines some curves with the topology of the given torus knot. 
Since the magnetic and electric helicities of these configurations are average measures of the linking 
number of the field lines and they are equal and conserved, it can be argued that some kind of nontrivial 
topology of these solutions can be found at any time. This kind of configurations might play an important role 
in different areas of research. 

If one were to ask the question what light is there can be no simple answer. A consistent and unambiguous
theoretical explanation of all optical phenomena is furnished jointly by Maxwell's electromagnetic theory
and quantum theory. However, the ultimate nature of light have still some unanswered questions such as its 
topological properties. Solutions as the ones presented here may help to understand part of these questions.      

\section{Acknowledgement}
We thank Friedrich W. Hehl, Jos\'e Mar\'{\i}a Montesinos and Antonio F. Ra\~nada for discussions 
and encouragement. This work has been supported by the Spanish Ministerio de Ciencia e Innovaci\'on under project 
AYA2009-14027-C07-04.

\end{document}